\documentclass[10pt,journal]{IEEEtran}

\ifCLASSOPTIONcompsoc
  \usepackage[nocompress]{cite}
\else
  \usepackage{cite}
\fi
\ifCLASSINFOpdf
\else
\fi
\usepackage{amsmath}
\usepackage[dvipsnames]{xcolor}
\usepackage{graphicx}
\usepackage{dblfloatfix}
\usepackage{amssymb, pifont, makecell}
\newcommand{\cmark}{\textcolor{Violet}{\ding{51}}} 
\newcommand{\ocircle}{\textcolor{ForestGreen}{\textcircled{}}} 
\usepackage{booktabs}
\usepackage[table]{xcolor}
\usepackage{tabularx}
\renewcommand{\arraystretch}{1.22}
\usepackage{colortbl}
\usepackage{array}
\usepackage{fixltx2e}
\usepackage{csquotes}
\usepackage{textcomp}
\usepackage{hyperref}

\begin{document}
%
\title{The Role of ISAC in 6G Networks: Enabling Next-Generation Wireless Systems}
%
%
%
%

\author{Muhammad Umar Farooq Qaisar, ~\IEEEmembership{Member,~IEEE,} Weijie Yuan, ~\IEEEmembership{Senior Member,~IEEE,} Onur Günlü, ~\IEEEmembership{Senior Member,~IEEE,} Taneli Riihonen, ~\IEEEmembership{Senior Member,~IEEE,} Yuanhao Cui, ~\IEEEmembership{Member,~IEEE,} Lin Zhang, ~\IEEEmembership{Senior Member,~IEEE,} Nuria Gonzalez-Prelcic, ~\IEEEmembership{Fellow,~IEEE,} Marco Di Renzo, ~\IEEEmembership{Fellow,~IEEE,} and Zhu Han, ~\IEEEmembership{Fellow,~IEEE}
\IEEEcompsocitemizethanks{
\IEEEcompsocthanksitem M. U. F. Qaisar (muhammad@buaa.edu.cn) is with Hangzhou International Innovation Institute of Beihang University, Hangzhou 311115, China;\protect
\IEEEcompsocthanksitem W. Yuan (yuanwj@sustech.edu.cn) is with the School of Automation and Intelligent Manufacturing, Southern University of Science and Technology, Shenzhen, China;\protect
\IEEEcompsocthanksitem O. Günlü (onur.guenlue@tu-dortmund.de) is with the Lehrstuhl für Nachrichtentechnik, Technische Universität Dortmund, Germany and Information Theory and Security Laboratory, Linköping University, Sweden;\protect
\IEEEcompsocthanksitem T. Riihonen (taneli.riihonen@tuni.fi)  is with the Faculty of Information Technology and Communication Sciences, Tampere University, Korkeakoulunkatu 1, 33720 Tampere, Finland;\protect
\IEEEcompsocthanksitem Y. Cui (yuanhao.cui@bupt.edu.cn) is with Beijing University of Posts and Telecommunications (BUPT), Beijing, China;\protect
\IEEEcompsocthanksitem L. Zhang (zhanglin@buaa.edu.cn) is with Hangzhou International Innovation Institute of Beihang University, Hangzhou 311115, China, the School of Automation Science and Electrical Engineering at Beihang University, Beijing 100191, China, and the State Key Laboratory of Intelligent Manufacturing Systems Technology, Beijing 100854, China;\protect
\IEEEcompsocthanksitem N. Gonzalez-Prelcic (ngprelcic@ucsd.edu) is with the Department of Electrical and Computer Engineering, University of California, San Diego, La Jolla, 92093 CA USA;\protect
\IEEEcompsocthanksitem M. Di. Renzo is with Universit\'e Paris-Saclay, CNRS, CentraleSup\'elec, Laboratoire des Signaux et Syst\`emes, 3 Rue Joliot-Curie, 91192 Gif-sur-Yvette, France. (marco.di-renzo@universite-paris-saclay.fr), and with King's College London, Centre for Telecommunications Research -- Department of Engineering, WC2R 2LS London, United Kingdom (marco.di\_renzo@kcl.ac.uk);\protect
\IEEEcompsocthanksitem Z. Han (hanzhu22@gmail.com) is with the Department of Electrical and Computer Engineering, University of Houston, Houston, TX 77004 USA;\protect
\IEEEcompsocthanksitem Corresponding author: Muhammad Umar Farooq Qaisar (muhammad@buaa.edu.cn);\protect
}}

%
%

\markboth{
}%
{Shell \MakeLowercase{\textit{et al.}}: Bare Advanced Demo of IEEEtran.cls for IEEE Computer Society Journals}
%



\IEEEtitleabstractindextext{%
\begin{abstract}
The commencement of the sixth-generation (6G) wireless networks represents a fundamental shift in the integration of communication and sensing technologies to support next-generation applications. Integrated sensing and communication (ISAC) is a key concept in this evolution, enabling end-to-end support for both communication and sensing within a unified framework. It enhances spectrum efficiency, reduces latency, and supports diverse use cases, including smart cities, autonomous systems, and perceptive environments. This tutorial provides a comprehensive overview of ISAC's role in 6G networks, beginning with its evolution since 5G and the technical drivers behind its adoption. Core principles and system variations of ISAC are introduced, followed by an in-depth discussion of the enabling technologies that facilitate its practical deployment. The paper further analyzes current research directions to highlight key challenges, open issues, and emerging trends. Design insights and recommendations are also presented to support future development and implementation. This work ultimately tries to address three central questions: Why is ISAC essential for 6G? What innovations does it bring? How will it shape the future of wireless communication?
\end{abstract}

}

\maketitle

\IEEEdisplaynontitleabstractindextext

%
\IEEEpeerreviewmaketitle

\ifCLASSOPTIONcompsoc
\IEEEraisesectionheading{\section{Introduction}\label{sec:introduction}}
\else
\section{Introduction and Background} \label{IB}
\label{sec:introduction}
\fi

%
%
%
%
\IEEEPARstart{T}{he} importance of integrated sensing and communication (ISAC) in the sixth-generation (6G) mobile networks is increasingly acknowledged as a crucial service for beyond-5G (B5G) technology \cite{1.0, 1.1, 1.2, 3.4.2.1}. Conventional networks have treated sensing (radar, localization, environmental monitoring) and data transmission as distinct operations, but ISAC amalgamates them into a unified radio-frequency (RF) front-end and a unique waveform.  In addition to offering significant hardware and cost benefits, ISAC's unification of diverse functions under a cohesive framework establishes the foundation for a new category of applications in 6G and beyond \cite{1.3, 1.4, 1.5, 4.3.1}.

\begin{figure}[t]
	\includegraphics[width=\linewidth,height=0.25\textheight]{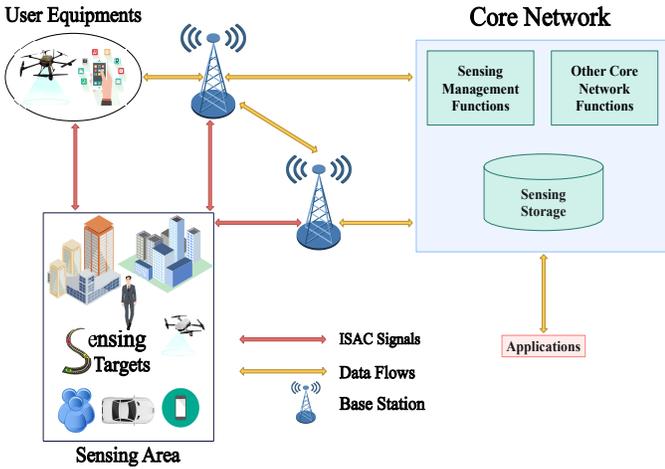}
	\centering
	\caption{Networked ISAC architecture demonstrating distributed sensing and communication.}
	\label{fig:ISAC-Network-Architecture}
\end{figure}

The International Telecommunication Union (ITU) has formally acknowledged ISAC as one of the six important 6G use cases, highlighting its significance in the development of next-generation wireless networks \cite{1.6},\cite{4.2.1}. The acknowledgment stems from ISAC's potential to deliver seamless and comprehensive intelligence across diverse sectors, including Industry 4.0 automation, smart cities, immersive extended-reality platforms, and autonomous vehicle networks. ISAC facilitates autonomous systems in sensing their environment with the identical waveform employed for communication, hence strengthening situational awareness in real time while reducing latency and spectrum overhead. ISAC significantly enhances spectrum efficiency by facilitating the shared use of spectrum, baseband hardware, RF front ends, and signal processing units. This is extremely significant as the sub-6 GHz and mmWave frequencies are increasingly congested. Spectrum reuse is particularly effective in scenarios like intelligent transportation and the coordination of unmanned aerial vehicles (UAVs), where simultaneous communication and environmental awareness are essential on a broad scale. ISAC also establishes a technological basis for networked sensing, allowing wireless infrastructure, such as base stations, to function as sensory nodes that collectively broadcast data and collect ambient information, as illustrated in Fig.~\ref{fig:ISAC-Network-Architecture}. This architecture enhances network-wide functions such as collaborative environmental mapping, instantaneous trajectory forecasting, and human activity identification \cite{1.777, 1.8, 1.9, 1.10}.

Within the framework of the internet of things (IoT) ecosystems, ISAC enables dispersed objects to sense and communicate with their environment. As IoT nodes take on ever greater responsibility for operations like resource allocation, autonomous navigation, and cooperative decision-making, it is essential that they are able to sense and communicate in a unified way \cite{1.11, 1.12, 1.13}. In contrast to conventional systems that operate in a sense-then-communicate structure, ISAC enables sensing-with-communication, which lowers processing overhead and latency, both of which are critical for making real-time decisions. Furthermore, ISAC offers significant potential for resolving issues with intelligent transportation systems. ISAC facilitates the simultaneous use of the same spectrum for radar sensing and communication in applications such as collaborative UAV navigation and predictive beamforming for connected vehicles.  As a result, beamforming delays and signaling overhead are reduced, and vehicular and aerial networks become more reliable \cite{1.14, 1.15, 1.16}. Critically, in dense urban deployments where communication channels must be integrated with safety-critical sensing capabilities, ISAC offers an appropriate response to the increasing demand for high-precision localization and environmental perception. In addition to improving communication resilience, ISAC facilitates high-resolution mapping of dynamic environments through techniques like multistatic sensing, channel state information (CSI)-assisted inference, and beam tracking \cite{1.17, 1.18, 1.19, 5.2.2}.

Numerous research studies indicate that ISAC is crucial for the future architecture of networks, especially concerning networked sensing and distributed intelligence \cite{3.4.1, 4.3.2, 1.20, 1.21, 1.22, 3.3.3}. Cell-free massive multiple input multiple output (MIMO) and cloud radio access network (C-RAN) systems can leverage ISAC-facilitated nodes that perform both communication tasks and collaborative sensing to interpret intricate environments \cite{3.4.2.1}, \cite{1.23, 1.24, 1.25, 1.26, 1.27, 6.3.10}.

\subsection{Background on 6G and Perceptive Networks}
The vision for 6G wireless networks anticipates unprecedented capabilities, including terabit-per-second peak data rates, centimeter-level positioning accuracy, and sub-millisecond radio interface latency. Attaining these ambitious objectives is impractical within the conventional \enquote{communication-only} framework described by 5G new radio (NR) \cite{1.28, 1.29, 6.3.7}. Conversely, 6G needs a paradigmatically new network structure that proactively perceives its radio and physical environments, thus turning the network into a sensory system. A 6G base station will not just transmit data frames but also inspect the echoes of these transmissions to create real-time environmental maps. This encompasses detection and localization of users, reflectors, and obstacles in the radio environment. These sensing features provide advanced functionalities, including adaptive beam alignment, blockage prediction, and proactive resource scheduling, which are critical in high-mobility environments such as high-speed rail and urban air mobility \cite{1.30, 1.31, 1.32}.

This integration is at the core of the 6G RAN architecture, which forms the foundation for the ISAC concept. By combining radar sensing and communication technologies, ISAC leverages the growing similarities in hardware architectures, channel characteristics, and signal processing methods, especially as both domains advance toward higher frequency bands and larger antenna arrays. It attempts to integrate sensing and communication functionality so that they can share resources and mutually support each other's performance. This provides superior spectral and energy efficiency, reduced hardware and signaling costs, and enables ubiquitous sensing services. These services are imperative in the development of future intelligence in smart environments, enabling applications such as vehicle to everything (V2X) communication, smart home, smart factory, remote sensing, environmental monitoring, and human-computer interaction \cite{2.2.2a}\cite{2.2.2},\cite{3.4.1},\cite{3.4.2.1}.

The integration of radar and communication systems has evolved over decades. Early radar systems, such as the phased-array radar developed during World War II, inspired the development of multi-antenna communication systems like MIMO, which became foundational in 3G to 5G networks\cite{1.33, 1.34}. Conversely, MIMO communication techniques influenced the design of advanced radar systems, such as co-located MIMO radar, enhancing sensing capabilities \cite{1.35}. In recent years, the convergence has deepened with the emergence of ISAC research, motivated by programs like the US Office of Naval Research’s Advanced Multifunction Radio Frequency Concept, which sought to integrate radar, communication, and electronic warfare functions into a common platform \cite{2.2.2},\cite{4.3.1}. ISAC research has explored embedding communication information in radar waveforms and leveraging communication-centric waveforms like orthogonal frequency division multiplexing (OFDM) for sensing purposes. The progression of massive MIMO and millimeter-wave technologies has further facilitated the integration of sensing and communication. Massive MIMO arrays, enabled by millimeter-wave (mmWave) frequencies, provide high beamforming gains and compact antenna arrays, but pose challenges in hardware cost and energy consumption \cite{1.36},\cite{5.3.1}. Hybrid analog-digital architectures have been proposed to address these challenges \cite{1.37, 1.38, 1.39}, similar to those mirrored in phased-MIMO radar designs that balance between phased array and MIMO radar benefits \cite{1.40},\cite{1.41}.

\begin{figure*}[t]
	\includegraphics[width=\linewidth]{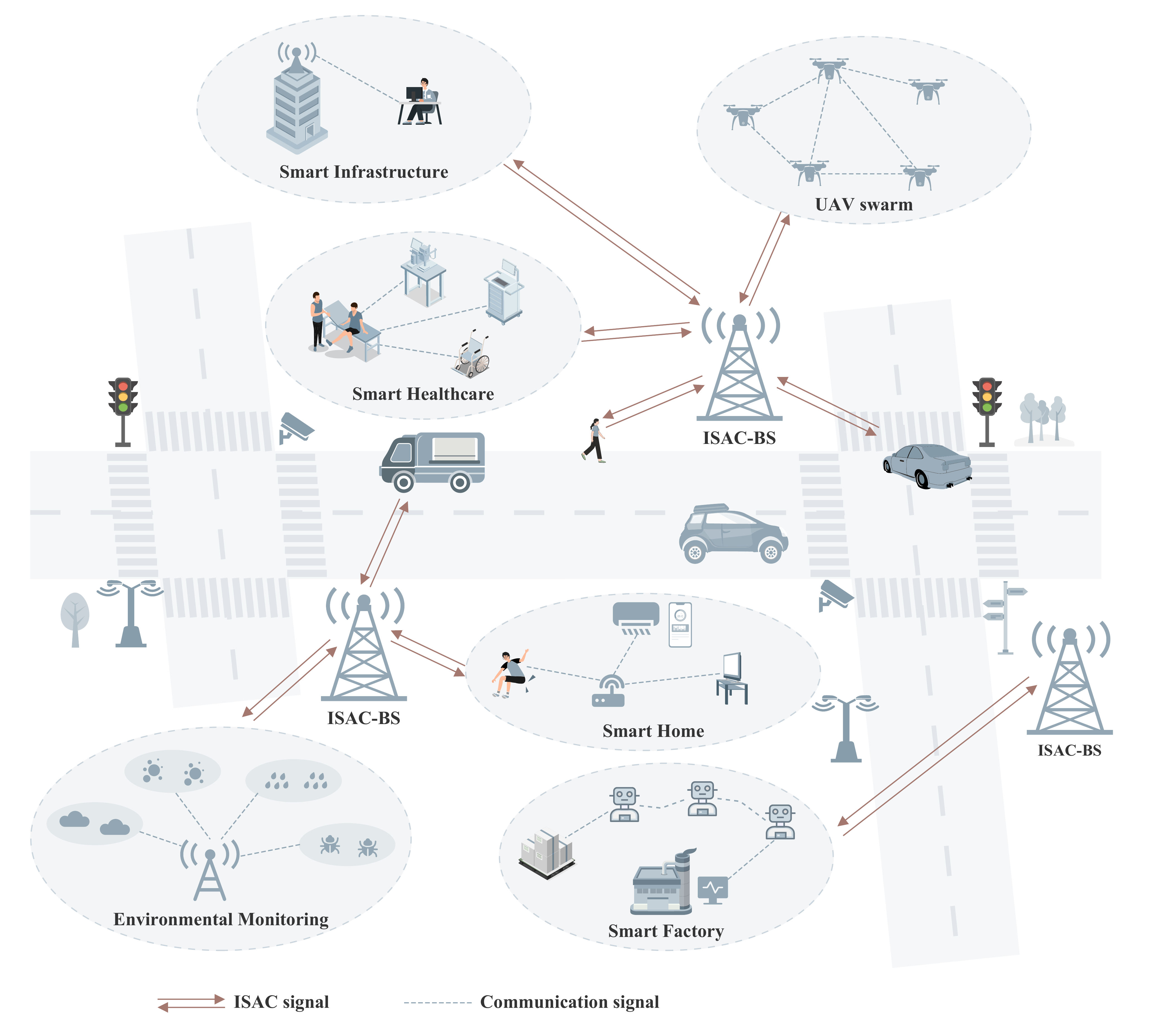}
	\centering
	\caption{Conceptual overview of ISAC-enabled 6G networks.}
	\label{fig:ISAC-enabled-6G}
\end{figure*}

\subsection{Tutorial Objectives, Audience, and Scope}
In this tutorial, a comprehensive and integrative overview of integrated sensing and communication technologies is presented, spanning from foundational theory to practical deployment considerations. The overview begins by tracing the evolution from early radar-communication coexistence approaches to modern dual-functional transceivers that jointly design sensing and communication waveforms. This historical perspective highlights the key motivations and milestones that have driven the convergence of these two traditionally separate domains, ultimately leading to modern networked ISAC architectures where base stations and distributed nodes jointly support both sensing and communication functions, as illustrated in Fig.~\ref{fig:ISAC-enabled-6G}.

\begin{figure*}[t]
	\includegraphics[width=\linewidth,height=\linewidth]{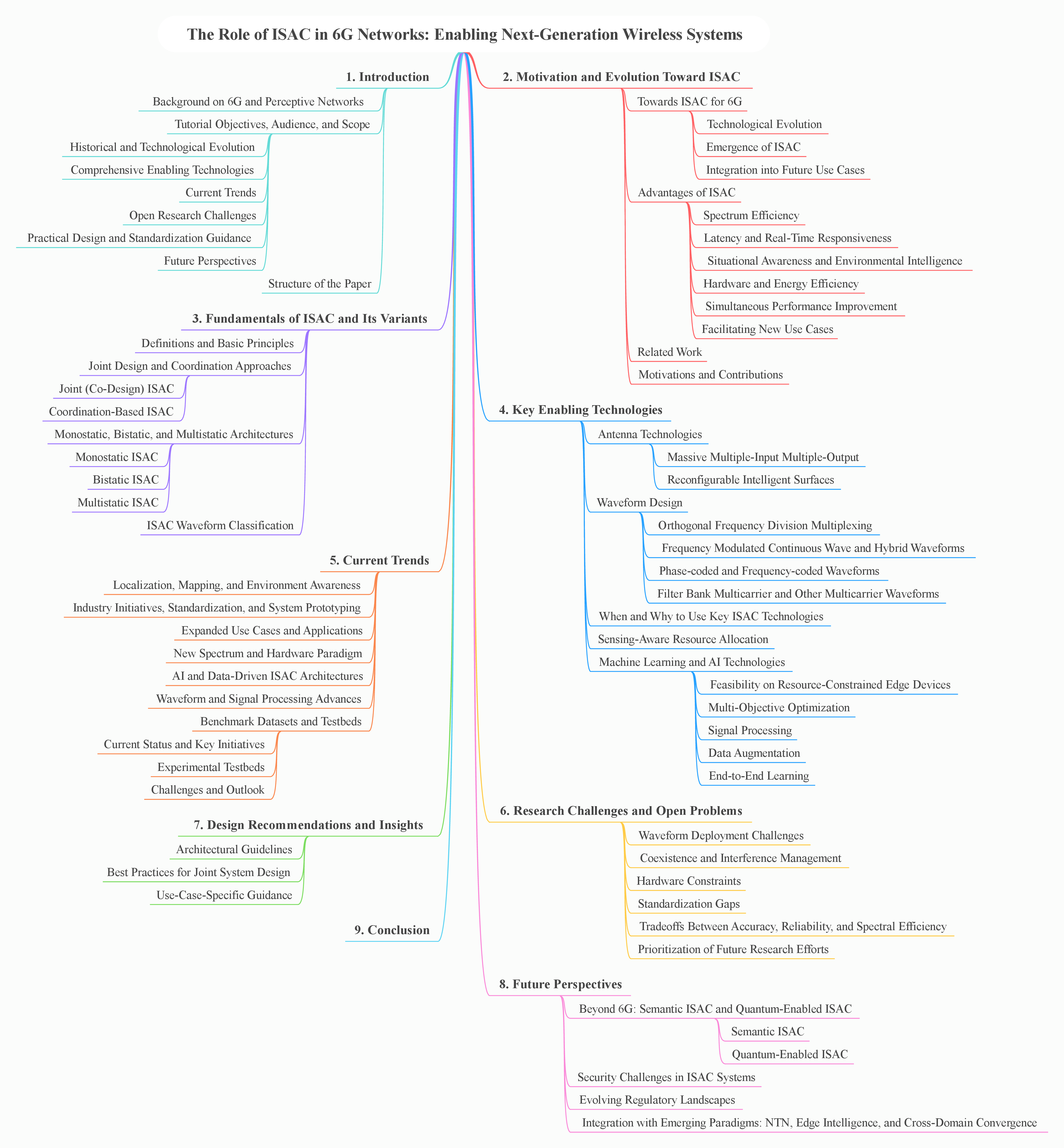}
	\centering
	\caption{Structural flow of the tutorial with layered sections and corresponding sub-sections}
	\label{fig:Main}
\end{figure*}

\begin{table*}[ht]
	\caption{List of Acronyms}
	\label{tab:acronyms}
	\centering
	\begin{tabular}{|p{0.10\textwidth}|p{0.33\textwidth}|p{0.10\textwidth}|p{0.35\textwidth}|}
		\hline
		\rowcolor[HTML]{E6E6E6}
		\textbf{Acronym} & \textbf{Description} & \textbf{Acronym} & \textbf{Description} \\
		\hline
		\hline
		ISAC  & Integrated Sensing And Communication  
		& 6G    & Sixth-Generation \\
		B5G   & Beyond 5G  
		& RF    & Radio-Frequency \\
		ITU   & International Telecommunication Union  
		& UAVs  & Unmanned Aerial Vehicles \\
		IoT   & Internet-of-Things  
		& CSI   & Channel State Information \\
		MIMO  & Multiple Input Multiple Output  
		& C-RAN & Cloud Radio Access Network \\
		NR    & New Radio  
		& V2X   & Vehicle To Everything \\
		OFDM  & Orthogonal Frequency Division Multiplexing  
		& mmWave& Millimeter-Wave \\
		RIS   & Reconfigurable Intelligent Surface  
		& THz   & Terahertz \\
		AI    & Artificial Intelligence  
		& URLLC & Ultra-Reliable Low-Latency Communication \\
		XR    & Extended Reality  
		& RCC   & Radar-Communication Coexistence \\
		NLOS  & Non-Line-Of-Sight  
		& ISCC  & Integrated Sensing, Communication, And Computation \\
		ISC   & Integrated Sensing And Computation  
		& ICC   & Integrated Communication And Computation \\
		3GPP  & 3rd Generation Partnership Project  
		& IEEE  & Institute Of Electrical And Electronics Engineers \\
		LFM   & Linear Frequency Modulation  
		& FBMC  & Filter Bank Multicarrier \\
		OTFS  & Orthogonal Time Frequency Space  
		& IIoT  & Industrial Internet Of Things \\
		FIM   & Fisher Information Matrix  
		& CRB   & Cramér-Rao Bound \\
		CP    & Cyclic-Prefix  
		& FFT   & Fast Fourier Transform \\
		P-ACF & Periodic Autocorrelation Function  
		& ISL   & Integrated Sidelobe Level \\
		QAM   & Quadrature Amplitude Modulation  
		& PSK   & Phase-Shift Keying \\
		CD    & Code-Division  
		& CDMA  & Code-Division Multiple Access \\
		SC    & Single-Carrier  
		& PAPR  & Peak-To-Average Power Ratio \\
		FMCW  & Frequency-Modulated Continuous Wave  
		& SNR   & Signal-To-Noise Ratio \\
		LPWAN & Low Power Wide Area Network  
		& QoS   & Quality Of Service \\
		MMSE  & Minimum Mean Square Error  
		& IRS   & Intelligent Reflecting Surfaces \\
		STAR  & Simultaneously Transmitting And Reflecting  
		& DFT   & Discrete Fourier Transform \\
		IM    & Index Modulation  
		& S-IM  & Superposed Index Modulated \\
		DM    & Direction Modulation  
		& OQAM  & Offset Quadrature Amplitude Modulation \\
		UFMC  & Universal Filtered Multicarrier  
		& ML    & Machine Learning \\
		GANs  & Generative Adversarial Networks  
		& RL    & Reinforcement Learning \\
		SLAM  & Simultaneous Localization And Mapping  
		& LIDAR & Light Detection And Ranging \\
		ETSI  & European Telecommunications Standards Institute  
		& ISG   & Industry Specification Group \\
		KPI   & Key Performance Indicator  
		& IA    & Infrastructure Association \\
		SDRs  & Software-Defined Radios  
		& ITS   & Intelligent Transportation Systems \\
		NEW   & Next-Evolution Waveforms  
		& ODDM  & Orthogonal Delay-Doppler Division Multiplexing \\
		AFDM  & Affine Frequency Division Multiplexing  
		& DDOP  & Delay-Doppler Orthogonal Pulse \\
		BER   & Bit Error Rate  
		& NTN   & Non-Terrestrial Networks \\
		MBSE  & Model-Based Systems Engineering
		& PSL   & Physical Layer Security \\
		SIC   & Self-Interference Cancellation
		& DSS   & Dynamic Spectrum Sharing \\
		WSN   & Wireless Sensor Networks 
		& SDN   & Software Defined Networking \\
		\hline
	\end{tabular}
\end{table*}

\subsubsection{Historical and Technological Evolution} 
We trace the progression from early radar-communication coexistence paradigms to the modern era of dual-functional, joint-design ISAC transceivers. This historical context elucidates how sensing and communication technologies have gradually converged, highlighting the drivers and milestones that have shaped the ISAC landscape.
\subsubsection{Comprehensive Enabling Technologies} 
We systematically catalog critical enabling technologies underpinning ISAC, including advanced antenna architectures (e.g., massive MIMO and reconfigurable intelligent surface (RIS)), wideband and terahertz (THz) waveform designs, machine learning assisted signal processing, and networked sensing infrastructures. This broad coverage bridges the gap between component-level innovations and system-level integration.
\subsubsection{Current Trends} 
We spotlight the latest advancements in ISAC, encompassing localization, mapping, and environment-aware sensing, industry-driven standardization and large-scale prototyping, artificial intelligence (AI)-empowered joint design, advances in mmWave/THz spectrum and hardware paradigms, breakthroughs in waveform and signal processing, expanded intelligent applications across transportation and industry, the emergence of data-driven architectures, and the development of benchmark datasets and testbeds.
\subsubsection{Open Research Challenges}
Building on a structured “Theory–System–Network–Application” narrative, we articulate the most pressing open problems across multiple layers from fundamental information-theoretic limits and waveform optimization to system design and large-scale network coordination. This roadmap identifies both well-explored areas and nascent challenges that require further investigation to propel ISAC towards widespread adoption.
\subsubsection{Practical Design and Standardization Guidance}
Recognizing the critical role of real-world deployments, we provide actionable design insights and practical guidelines for engineers and researchers preparing for field trials and contributing to standardization efforts. This emphasis facilitates the translation of ISAC concepts from theory to practice, supporting the evolution of 6G and beyond.
\subsubsection{Future Perspectives}
We project the anticipated evolution of ISAC beyond 6G, highlighting trends towards semantic and quantum-enabled networks, integration with non-terrestrial and edge-intelligent systems, and the convergence of perceptive wireless ecosystems driven by regulatory, security, and cross-domain collaboration initiatives.

This tutorial delivers a comprehensive and focused overview of ISAC, emphasizing the synergy and trade-offs between sensing and communication in next-generation wireless networks. It explores the unification of radar sensing and communication systems through both waveform design and hardware innovations that enable efficient dual-functional transceivers. By synthesizing recent advancements across waveform strategies, antenna technologies, signal processing, network architectures, and practical deployments, the tutorial offers a cohesive framework that bridges gaps left by prior works concentrating on particular facets rather than a comprehensive perspective. It also highlights emerging trends such as machine learning integration, terahertz-band ISAC, and networked sensing backhaul, while providing a structured classification of open research challenges aligned with the evolving wireless ecosystem. Addressing both standardization efforts and real-world deployment considerations, this work serves as a valuable and timely reference for advancing ISAC technologies toward practical implementation.

\subsection{Structure of the Paper}
In this paper, we provide a comprehensive tutorial on ISAC for $6G$ networks as shown in Fig.~\ref{fig:Main} and Table~\ref{tab:acronyms}
lists the acronyms used. We begin with the introduction in Section~\ref{IB}, establishing ISAC's foundational context by tracing the historical evolution from radar-communication coexistence to modern 6G perceptive networks while defining tutorial objectives, scope, and structural organization. Section~\ref{METI} presents the technical foundations of ISAC by examining the transition toward 6G dual-functional architectures, analyzing key advantages, reviewing current research landscape, and articulating core motivations driving ISAC development. Section~\ref{FIIV} presents the fundamentals of ISAC and its variants, covering key definitions, principles, and system types. Section~\ref{KET} discusses the key enabling technologies, including advanced antennas, waveform design, resource allocation, and AI integration. In Section~\ref{CT}, we summarize current trends, industry initiatives, and notable academic work shaping the ISAC landscape. Section~\ref{RCOP} identifies the main research challenges and open problems, such as interference management, hardware constraints, and standardization issues. Section~\ref{DRI} offers practical design recommendations and insights for ISAC deployment. Section~\ref{FP} explores future perspectives, including the evolution of ISAC beyond $6G$ and integration with other emerging paradigms. Finally, Section~\ref{conclude} concludes with a summary of key points and the outlook for ISAC in next-generation wireless systems.

\section{Motivation and Evolution Toward ISAC}\label{METI}

\subsection{Towards ISAC for 6G}
The transition from 5G to 6G would imply a deep transformation of wireless network architecture and characteristics, where ISAC is regarded as an essential platform of future networks \cite{1.1},\cite{6.3.7}. While 5G has introduced unparalleled communication technologies, such as ultra-reliable low-latency communication (URLLC), mmWave access, and massive MIMO, these achievements are more about communication \cite{1.1.1},\cite{1.1.2}. Conversely, 6G perceives a bi-functional architecture integrating communication and high-accuracy sensing to enable the network to be perceptive and context-aware. This is being driven by growing needs for smart infrastructure, autonomous mobility, human-machine interaction, and extended reality (XR) applications, which need real-time situational awareness and environmental perception \cite{1.1.3},\cite{4.6.3}.

\subsubsection{Technological Evolution}
5G networks feature URLLC, mmWave communications, and massive MIMO, establishing a foundation for higher data rates and device connections. Nonetheless, these advancements are solely inadequate to achieve the enormous 6G goals, including terabit-per-second peak rates, centimeter-level positioning precision, and sub-millisecond latency. The advances in hardware and signal processing for 6G, including elevated frequency ranges (notably terahertz), smaller devices, and larger antenna arrays, inherently integrate sensing and communication technologies, rendering ISAC a practical and crucial evolution. The development of radar and communication systems has historically influenced one another. Initial phased-array radars influenced multi-antenna MIMO communication systems, which subsequently resulted in concepts for MIMO radars that make use of spatial degrees of freedom and waveform diversity. This mutual inspiration culminates in ISAC, where sensing and communication functions are no longer separate but integrated within a unified transceiver design. This integration promises spectral and energy efficiency gains, cost reductions, and enhanced performance through mutual assistance; communication can aid sensing accuracy, and sensing can improve communication reliability \cite{1.1.1.1, 1.1.1.2, 1.1.1.3}.

\subsubsection{Emergence of ISAC}
The technology moves beyond mere coexistence or spectral sharing of radar and communication systems, aiming instead for a deep integration where both functions share hardware, waveforms, and signal processing. This approach enables the network to become perceptive, capable of real-time environmental mapping and adaptive resource management \cite{1.15},\cite{4.3.2}. The concept of ISAC has evolved through several phases: i) \textit{Radar-Communication Coexistence (RCC)}: Early efforts focused on managing interference when radar and communication systems operate in overlapping frequency bands. ii) \textit{Joint Waveform Design}: Embedding communication information into radar waveforms (e.g., chirp signals with phase-shift keying) and using communication waveforms (e.g., OFDM) for sensing. iii) \textit{Dual-Functional Transceivers}: Development of hardware architectures such as hybrid analog-digital arrays that support both sensing and communication efficiently \cite{1.1.1.4},\cite{1.1.1.5},\cite{2.2.2}.

\subsubsection{Integration into Future Use Cases}
ISAC is envisioned to underpin a variety of emerging 6G use cases requiring simultaneous sensing and communication, including i) \textit{High-Mobility Scenarios:} Such as high-speed rail and urban air mobility, where real-time environmental awareness is critical for safety and connectivity. ii) \textit{Smart Environments:} Like smart homes, smart industry, and green monitoring, where sensor information enriches the communication services in enabling automation and context awareness. iii) \textit{Vehicle to Everything Communications:} Such as integrated sensing, enhancing communication reliability, and situational awareness for autonomous vehicle operation. iv) \textit{Human-Machine Interaction:} ISAC facilitates real-time posture assessment, gesture identification, and complete immersion by incorporating sensor capabilities within the communication layer \cite{1.1.1.6, 2.2.4, 6.3.8, 7.2.1}.

Beyond specific applications, ISAC fundamentally reshapes the architecture of the IoT \cite{IoT} and wireless sensor networks (WSN) \cite{TORAS}. By integrating sensing into massive IoT deployments, ISAC eliminates dedicated sensor hardware, while software-defined networking (SDN) enables the centralized orchestration of these nodes, allowing the network to dynamically shift between communication and distributed sensing modes based on real-time demands \cite{load, PoisedG, SDN-WRSN}.

ISAC, through the establishment of a perceptive network, enhances proactive resource scheduling, blockage prediction, and adaptive beamforming, essential for stable 6G networks, and simultaneously decreases infrastructure expense and spectrum usage, independent of hardware complexity and energy efficiency issues.

\subsection{Advantages of ISAC}
ISAC brings numerous advantages that are at the core of future wireless network vision, especially for 6G and beyond. The advantages arise through sensing and communication functions being integrated together, previously separately designed and implemented. ISAC's main advantages are described in the following sections in terms of spectrum efficiency, latency, situational awareness and environmental intelligence, hardware and energy efficiency, and shared performance improvement \cite{1.21},\cite{1.2.1},\cite{3.4.2}.

\subsubsection{Spectrum Efficiency}
ISAC considerably improves spectrum efficiency by enabling the simultaneous use of communication and sensing within the same frequency resources. Communication and radar typically contend for spectrum in conventional systems, causing scattered allocations and suboptimal use. By integrating these capabilities into a single platform, ISAC removes the necessity of separate spectral bands for each process. Sophisticated waveform forms, such as OFDM and chirp signals, enable the concurrent extraction of environmental information and transmission of communications data. Effective spectrum sharing becomes ever more important as networks shift to higher frequency bands, such as terahertz and millimeter-wave, where bandwidth is in high demand and resources are scarce \cite{1.2.1.1},\cite{1.2.1.2}.

\subsubsection{Latency and Real-Time Responsiveness}
Sensing and communication integration minimizes system latency via real-time monitoring and data sharing in an integrated environment. Decoupled structures of conventional systems, separating radar and communication, introduce delays incurred by cross-system coordination. ISAC, by contrast, allows base stations to directly interpret echoes from transmitted signals, enabling real-time mapping of reflectors, users, and obstacles. This real-time feature enables sub-millisecond radio-interface latency, which is essential for applications that include immersive augmented reality, autonomous transportation, and industrial automation. The network's capability to offer immediate situational information enables efficient decision-making and response \cite{1.2.2.1},\cite{4.2.2}.

\subsubsection{Situational Awareness and Environmental Intelligence}
In order to provide wireless networks with constant situational awareness, ISAC leverages the communication signal echoes to facilitate ambient sensing. This enables the construction of real-time maps of the network's surroundings based on the detection and tracking of moving objects (e.g., drones and vehicles) and object velocity and trajectory estimation. Environmental intelligence of this sort is instrumental to advanced functions like preemptive resource allocation, blockage prediction, and adaptive beam alignment. The network capability to collect and process sensory information enables learning and intelligence growth, enabling a large range of environment-aware and location-aware use cases. This is especially beneficial in high-mobility environments, such as urban air mobility and high-speed trains, where environmental conditions vary rapidly and must be adapted quickly \cite{1.2.3.1, 1.2.3.2, 4.5.3}.

\subsubsection{Hardware and Energy Efficiency}
ISAC presents substantial benefits in terms of hardware and energy efficiency by merging the conventional distinct communication and sensing functions into a single system. This synthesis enables the reuse of signal processing units, antennas, and RF chains, thereby decreasing the overall hardware complexity and the number of physical elements. Energy-efficient, shared infrastructure reduces duplicate signal transmissions and hardware activation, resulting in reduced power consumption in comparison to standalone sensing and communication units. It further reduces signaling overhead and allows for improved utilization of resources, hence found to be especially useful in crowded or energy-constrained environments such as autonomous systems and IoT. These developments foster a greener and economically sustainable model for wireless networks of the future \cite{5.2.2},\cite{1.2.4.1},\cite{3.1.2.10}.

\subsubsection{Simultaneous Performance Improvement}
Communication-aided sensing leverages network data to strengthen the accuracy and reliability of environmental perception, whereas sensing-aided communication leverages environmental data to strengthen link reliability, throughput, and resource allocation. This synergy translates into better overall system performance than for standalone or loosely coupled systems. ISAC's end-to-end model of integration ensures that the two functions are designed to complement one another in a reciprocal manner, as opposed to viewing the two as separate tasks \cite{1.2.5.1},\cite{1.2.5.2}.

\subsubsection{Facilitating New Use Cases}
Benefits of ISAC facilitate many new use cases that need both environmental awareness and connectivity. They are automated manufacturing, smart homes, vehicle-to-everything communications, advanced human-computer interaction, and environmental monitoring. It offers accurate environmental information and improved wireless connectivity, making it a key technology for the smart world of the future, facilitating ubiquitous context-aware applications and intelligent automation across industries \cite{1.1.1.6},\cite{1.2.6.1},\cite{1.2.6.2}.

\subsection{Related Work}
The landscape of ISAC for 6G encompasses multiple research threads, ranging from architectural innovations, signal design, learning algorithms, and security models, to integration with computation and application in emerging scenarios. Below, we summarize the research advancements, followed by a comparison highlighting how our tutorial distinctly covers and unifies the core discussions.

In \cite{3.1.2.9}, the authors presented a focused survey on metasurface-assisted ISAC, particularly involving RIS and holographic approaches, covering integration levels (coexistence and dual-function), deployment architectures, and applications such as beamforming, non-line-of-sight (NLOS) communication, interference management, and security, especially in automotive radar and vehicular networks. However, the survey is specialized and does not broadly address ISAC fundamentals beyond the metasurface context. In \cite{3.3.3}, the authors presented a broad survey on integrated sensing, communication, and computation (ISCC), covering the unification of ISAC, integrated sensing and computation (ISC), and integrated communication and computation (ICC), and addressing signal design, joint resource management, and applications such as digital twins, federated learning, smart cities, autonomous driving, and edge AI, with an emphasis on computational integration and task-oriented management. In \cite{3.4.2}, the authors presented a comprehensive review of learning algorithms, including machine learning and reinforcement learning, for ISAC, showing how data-driven methods address key tasks such as resource management, waveform and beamforming design, angle estimation, signal classification, and system security. They also discuss the challenges of traditional optimization versus learning-based approaches and examine applications in IoT, V2X, UAVs, human activity sensing, and wireless health. In \cite{3.4.1}, the authors presented a comprehensive survey of secure and intelligent ISAC, introducing an IoT architecture that combines ISAC with AI, edge/cloud computing, and security. The survey details system layers, hardware, intelligent middleware, and application, and outlines security and privacy requirements, performance metrics, and evaluation indexes for sensing, computation, and communication, with applications in smart health, smart factories, industrial IoT, disaster response, and precision agriculture.

In \cite{5.3.1}, the authors presented a thorough survey offering an evolutionary perspective on ISAC, covering developments from RF and optical technologies to single- and multi-cell, as well as multi-modal systems, with emphasis on standardization, security, edge computation, and dataset development. The survey addresses RF and optical ISAC, advances in collaborative architectures, and ongoing standardization efforts by 3rd generation partnership project (3GPP), ITU, and institute of electrical and electronics engineers (IEEE), with applications in IoT, vehicular, mobile, distributed networks, and perception-driven 6G. In \cite{4.6.3}, the authors conducted an in-depth survey of signal design and processing techniques for ISAC in 5G-Advanced and 6G from a mobile communication perspective, covering waveform optimization, ambiguity functions, Doppler sensitivity, and advanced waveforms like OFDM, linear frequency modulation (LFM), filter bank multicarrier (FBMC), and orthogonal time frequency space (OTFS), with emphasis on signal-level trade-offs and applications in intelligent mobility, localization, and dynamic spectrum sharing. In \cite{3.1.2.10}, the authors offered an extensive survey of recent advances in ISAC, focusing on physical layer and network design, and outlining ten open research challenges including information theory, beamforming, synchronization, Pareto-optimal signaling, and multi-task applications. The survey emphasizes integration and coordination gains, addressing unresolved issues such as super-resolution, resource management, and security in multi-object and multi-task contexts, with applications in IoT, V2X, human sensing, UAVs, smart homes, and industrial automation.

In \cite{2.2.2}, the authors outlined a full-scale survey of ISAC, covering advances in physical layer and network design, ten open research challenges, and key topics such as integration and coordination gains, super-resolution, resource management, and security for multi-object and multi-task scenarios. The review also addresses ISAC background, theory, applications, design, and network architectures, highlighting the dual roles of sensing and communication, a technology roadmap, and use cases including smart cities, localization, imaging, drone monitoring, and environmental mapping. In \cite{1.3.1}, the authors provided a detailed survey on ISAC and ISCC for the metaverse, covering metaverse platform architecture, enabling technologies like blockchain, edge intelligence, and 6G, and discussing applications in smart homes, smart factories, healthcare, transportation, UAVs, and space-air-ground integrated networks. In \cite{4.7.2.2}, the authors outlined a full-scale survey of six key integration strategies for ISAC in 6G, multi-node coordination, multiband operation, multimodal fusion, balancing model and data-driven approaches, programmable hardware, and network operation integration, highlighting a roadmap for next-generation ISAC with integration gains and illustrating applications in distributed sensing, multi-band networks, cross-layer optimization, and edge AI.

\vspace{1em}
\noindent\textbf{Distinctive Contribution: A \textquotedblleft Lifecycle-Based Integration\textquotedblright $\,$ Framework} 

While the aforementioned surveys provide valuable insights into specific ISAC subdomains, ranging from physical-layer signal processing \cite{4.6.3} and learning algorithms \cite{3.4.2} to security \cite{3.4.1}, metasurfaces \cite{3.1.2.9}, ISCC \cite{3.3.3}, and metaverse-oriented applications \cite{1.3.1}, they predominantly adopt a \emph{component-centric} view, treating waveforms, hardware, and networking as isolated verticals. As summarized in Table~\ref{tab:survey_comparison_revised}, this often results in a fragmented understanding of how physical-layer gains translate into system-level capabilities. Most prior works emphasize particular technology families or specific system layers and therefore offer limited critical commentary on why certain topics are treated only partially (marked as $\ocircle$ in Table~\ref{tab:survey_comparison_revised}) and how these omissions propagate along the theory, system, and standardization chain. A critical gap remains in offering a holistic framework that connects fundamental information-theoretic limits with practical deployment constraints and standardization.

To address this gap, our tutorial departs from the traditional feature-listing approach and introduces a \emph{\textquotedblleft Lifecycle-Based Integration\textquotedblright $\,$taxonomy}. We unify the disparate challenges of ISAC through a cohesive \emph{Theory-to-Standardization} lens, structured into three interconnected layers:
\begin{itemize}
	\item \textit{Fundamental Convergence Layer:} We unify the mathematical view by deriving a joint signal model (Section~\ref{FIIV}) that explicitly links communication throughput with sensing precision bounds (Cramér-Rao Bound (CRB)/Posterior Cramér-Rao Bounds (PCRB)), identifying the fundamental trade-offs often treated separately in prior works. This unified model serves as the foundation for all subsequent analysis and explains why fragmentation arises when these concepts are decoupled.
	\item \textit{Technological Realization Layer:} We analyze enabling technologies not as standalone tools, but as essential solvers for specific architectural bottlenecks, connecting massive MIMO, RIS, waveform families, sensing-aware resource allocation, and AI/ML directly to the practical non-idealities (e.g., synchronization, resource constraints, channel uncertainty) identified in the fundamental layer. This layer explicitly demonstrates why isolated technology surveys remain incomplete without integration to the signal model.
	\item \textit{Systemic Evolution Layer:} We extend the analysis beyond research concepts to the \emph{standardization lifecycle}, mapping theoretical innovations to concrete 3GPP Release 19/20 gaps and future roadmap milestones (Quantum/Semantic ISAC), along with industry initiatives and regulatory developments. This layer uniquely bridges the gap between academic research and practical deployment that most surveys do not adequately address.
\end{itemize}

Unlike surveys that concentrate on selected areas, such as metasurfaces, learning, ISCC, or metaverse applications, our tutorial bridges theory and application by connecting the historical context and technical breakthroughs to practical engineering guidance and deployment recommendations. Critically, this work goes beyond organizational breadth by providing a \emph{novel conceptual framework}, the Lifecycle-Based Integration approach, that explains \emph{why} certain gaps exist in prior works and \emph{how} those gaps are systematically closed. Section~\ref{FIIV} introduces a unified ISAC signal model and associated performance bounds (CRB/PCRB) that provide the mathematical foundation for the subsequent discussion. Building on these fundamentals, Section~\ref{KET} examines waveform design, resource allocation, and AI-assisted schemes in a manner consistent with this joint sensing and communication view. Sections~\ref{RCOP}--\ref{FP} leverage this same theoretical perspective to frame system-level challenges, design insights, and standardization directions. By anchoring the entire discussion in a common mathematical and conceptual foundation, rather than treating topics as isolated verticals, the tutorial ensures that technological advancements, architectural trade-offs, and standardization requirements are understood within a coherent, lifecycle-spanning framework.

\begin{table*}[htbp]
	\centering
	\caption{Comparative Coverage of ISAC Surveys and Tutorials Across Thematic Dimensions, Coverage Depth, and Recency}
	\renewcommand{\arraystretch}{1.2}
	\setlength{\tabcolsep}{3pt}
	\label{tab:survey_comparison_revised}
	\begin{tabular}{|c|c|c|c|c|c|c|c|c|c|p{8cm}|}
		\hline
		\rowcolor[HTML]{E6E6E6}
		\textbf{Ref.} & \textbf{Year} &
		\rotatebox{90}{\makecell{\textbf{Fundamentals/}\\\textbf{Architecture}}} & 
		\rotatebox{90}{\makecell{\textbf{Signal/}\\\textbf{Waveform}}} & 
		\rotatebox{90}{\makecell{\textbf{Resource Mgmt.}\\\textbf{/ AI}}} & 
		\rotatebox{90}{\makecell{\textbf{Security/}\\\textbf{Privacy}}} & 
		\rotatebox{90}{\makecell{\textbf{Learning}\\\textbf{Algorithms}}} & 
		\rotatebox{90}{\makecell{\textbf{Meta-/RIS/}\\\textbf{Imaging}}} & 
		\rotatebox{90}{\textbf{Standardization}} & 
		\rotatebox{90}{\makecell{\textbf{Application}\\\textbf{Scenarios}}} & 
		\makecell[c]{\textbf{Key Focus \& Limitation}} \\
		\hline
		\hline
		\cite{3.1.2.9} & 2024 & \ocircle & \ocircle & \ocircle & \ocircle & \ocircle & \cmark & \ocircle & \cmark & \textbf{Focus:} Metasurface-assisted ISAC. \newline \textbf{Limit:} Lacks broad system-level analysis beyond RIS context. \\
		\hline
		\cite{3.3.3} & 2024 & \cmark & \ocircle & \cmark & \ocircle & \ocircle & \ocircle & \ocircle & \cmark & \textbf{Focus:} ISCC (Comm-Sensing-Compute). \newline \textbf{Limit:} Less depth on physical layer signal models. \\
		\hline
		\cite{3.4.2} & 2025 & \ocircle & \cmark & \cmark & \cmark & \cmark & \ocircle & \ocircle & \cmark & \textbf{Focus:} Advanced ML algorithms. \newline \textbf{Limit:} Algorithm-centric, omits hardware/standardization gaps. \\
		\hline
		\cite{3.4.1} & 2025 & \cmark & \ocircle & \cmark & \cmark & \ocircle & \cmark & \ocircle & \cmark & \textbf{Focus:} Security \& Privacy. \newline \textbf{Limit:} Specialized on security, limited waveform design coverage. \\
		\hline
		\cite{5.3.1} & 2025 & \cmark & \cmark & \ocircle & \cmark & \ocircle & \cmark & \cmark & \cmark & \textbf{Focus:} Evolutionary perspective. \newline \textbf{Limit:} High-level overview, less detail on specific AI implementations. \\
		\hline
		\cite{4.6.3} & 2023 & \ocircle & \cmark & \ocircle & \ocircle & \ocircle & \ocircle & \ocircle & \cmark & \textbf{Focus:} Signal processing \& waveforms. \newline \textbf{Limit:} Lacks network-level protocol \& security coverage. \\
		\hline
		\cite{3.1.2.10} & 2024 & \cmark & \cmark & \cmark & \cmark & \ocircle & \ocircle & \ocircle & \cmark & \textbf{Focus:} Ten open challenges. \newline \textbf{Limit:} Broad scope but limited depth on emerging 6G paradigms. \\
		\hline
		\cite{2.2.2} & 2022 & \cmark & \cmark & \cmark & \ocircle & \ocircle & \ocircle & \ocircle & \cmark & \textbf{Focus:} Cellular sensing concepts. \newline \textbf{Limit:} Dated (2022), lacks recent 6G/AI advancements. \\
		\hline
		\cite{1.3.1} & 2024 & \ocircle & \ocircle & \cmark & \ocircle & \ocircle & \cmark & \ocircle & \cmark & \textbf{Focus:} Metaverse integration. \newline \textbf{Limit:} Application-specific, limited physical layer rigorousness. \\
		\hline
		\cite{4.7.2.2} & 2025 & \cmark & \cmark & \cmark & \ocircle & \ocircle & \cmark & \ocircle & \cmark & \textbf{Focus:} Integration avenues. \newline \textbf{Limit:} Forward-looking vision, less focus on current standards. \\
		\hline
		\textbf{Ours} & \textbf{2026} & \cmark & \cmark & \cmark & \cmark & \cmark & \cmark & \cmark & \cmark & \textbf{Differentiation:} Unifies theory, system design, and AI with practical \textbf{standardization} (3GPP Rel-19/20) \& \textbf{visionary roadmaps} (Quantum/Semantic). \\
		\hline
	\end{tabular}
	\vspace{2mm}
	
	\noindent
	\textbf{Legend:}~\cmark~Fully addressed, \ocircle~Partial/Contextual, \textbf{Focus/Limit}~Qualitative differentiation.
\end{table*}

\subsection{Motivations and Contributions}
The motivation for ISAC arises from growing requirements and technological advancements of future wireless networks.  As technologies progress towards 6G, there will be an increased necessity to facilitate both improved connection and accurate, fault-tolerant sensing within a cohesive architecture.  The two necessities arise from new applications of intelligent manufacturing, autonomous transportation, extended reality, and environmental monitoring that all depend on reliable communications and real-time situational awareness.

Technological improvements further justify ISAC. Sensing and communications are becoming more convergent in hardware structure, operating on higher frequency bands, employing larger arrays of antennas, and trending towards miniaturized technology. This convergence provides a unique opportunity to unify the two functionalities, allowing future networks to go beyond traditional communications and provide ubiquitous sensing services. Such capabilities enable networks to measure, image, and learn from their environments, laying the foundation for intelligence in the future smart world. The joint design of sensing and communication operations is thus essential to fully exploit the dense infrastructure of future networks and to construct perceptive, adaptive, and intelligent wireless systems.

This tutorial paper provides a comprehensive and up-to-date synthesis of ISAC in the context of 6G networks, offering several key contributions. First, we introduce a novel \emph{Lifecycle-Based Integration} framework that establishes a critical conceptual bridge missing from existing surveys. Unlike prior works that focus exclusively on physical layer algorithms or high-level use cases, this tutorial connects theoretical signal models directly to practical realization by unifying the ISAC landscape across three dimensions: theoretical limits (e.g., fundamental performance bounds in dynamic channels), hardware constraints (e.g., synchronization and jitter effects), and standardization bottlenecks. This distinctive lens provides the critical commentary absent in existing literature, explaining \emph{why} certain theoretical gains are difficult to realize in practice and \emph{how} standardization efforts address these gaps. Second, it traces the historical evolution and conceptual foundations of ISAC from early radar-communication coexistence to the present paradigm of dual-functional joint transceivers. Third, it systematically surveys enabling technologies such as advanced antenna architectures, waveform design, resource allocation, and machine learning integration. Fourth, the paper critically analyzes the fundamental tradeoffs and mutual benefits between sensing and communication. It reviews current trends, industry initiatives, and benchmark developments, identifies and articulates major open research challenges and standardization gaps, and delivers actionable design recommendations and forward-looking perspectives for deploying ISAC in next-generation wireless systems. By consolidating state-of-the-art knowledge and providing structured insights organized within the Lifecycle-Based Integration framework, this tutorial serves as a foundational reference for researchers, engineers, and standardization bodies, facilitating the advancement and practical realization of ISAC in 6G and beyond.

\textbf{\textit{Why now?}}
The urgency for ISAC arises from both the readiness of enabling technologies and the pressing demands of new applications. As 5G standardization solidifies, the wireless community is already looking ahead to what 6G will require. The convergence of sensing and communication hardware, driven by advances in massive MIMO, millimeter-wave, and terahertz technologies, makes it feasible to implement joint sensing and communication within a single, efficient system. This technological maturity coincides with the proliferation of applications that cannot be adequately served by communication-only networks, such as high-precision localization, real-time environmental mapping, and context-aware automation.

Furthermore, the historical separation of radar and communication systems has led to inefficiencies in spectrum usage, hardware redundancy, and increased operational costs. The ISAC paradigm addresses these challenges by integrating both functionalities, thereby improving spectral and energy efficiency, reducing hardware and signaling costs, and enabling mutual performance gains through communication-assisted sensing and sensing-assisted communication. The growing attention from both academia and industry, as well as ongoing standardization activities, underscores the timeliness and necessity of ISAC for the next generation of wireless networks.

\section{Fundamentals of ISAC and Its Variants} \label{FIIV}

\subsection{Definitions and Basic Principles} \label{subsec:DBP}
ISAC merges radar-style sensing and data transmission within a single radio front-end, waveform, and signal-processing chain. By doing so, it realizes an \emph{integration gain} through shared spectrum and hardware resources, and a \emph{coordination gain}, in which sensing and communication routines mutually enhance one another \cite{2.1.1, 2.1.2}. 

The fundamental ISAC signal model builds upon a unified wideband waveform, $\mathbf{x}(t) \in \mathbb{C}^{N_t \times 1}$, which the Base Station (BS) transmits to serve both communication and sensing tasks simultaneously. In a typical downlink monostatic scenario, this single transmission propagates through two distinct channels. First, the communication signal traverses a multipath environment to reach the User Equipment (UE). Second, the sensing signal reflects off environmental targets and returns to the BS. Consequently, the system is defined by two coupled reception models.

For the communication link, the signal received at the UE, denoted as $\mathbf{y}_c(t) \in \mathbb{C}^{N_r \times 1}$, is a superposition of $P$ multipath components. This relationship is expressed as:
\begin{equation} 
	\begin{split}
		\mathbf{y}_c(t) = & \sum_{p=1}^{P} \beta_p \mathbf{a}_{rx, UE}(\theta_{rx, p}) \mathbf{a}_{tx, BS}^T(\theta_{tx, p}) \\
		& \times \mathbf{x}(t - \tau_{c,p}) + \mathbf{n}_c(t),
	\end{split}
	\label{eq:comm_model}
\end{equation}

where $P$ is the number of communication paths, $\beta_p$ and $\tau_{c,p}$ are the complex channel gain and delay of the $p$-th path, and $\mathbf{a}_{rx, UE}$ and $\mathbf{a}_{tx, BS}$ denote the array response vectors for the UE receive and BS transmit antennas, respectively.

Simultaneously, the BS operates as a monostatic radar receiver, capturing echoes reflected from $L$ targets in the environment. The received sensing signal at the BS, $\mathbf{y}_s(t) \in \mathbb{C}^{N_t \times 1}$, is modeled as:
\begin{equation} 
	\begin{split}
		\mathbf{y}_s(t) = & \sum_{l=1}^{L} \alpha_l e^{j2\pi f_{D,l} t} \mathbf{a}_{rx, BS}(\phi_{rx, l}) \mathbf{a}_{tx, BS}^T(\phi_{tx, l}) \\
		& \times \mathbf{x}(t - \tau_{s,l}) + \mathbf{n}_s(t),
	\end{split}
	\label{eq:sensing_model}
\end{equation}
where $\alpha_l$, $f_{D,l}$, and $\tau_{s,l}$ represent the reflection coefficient, Doppler shift, and round-trip delay of the $l$-th target, respectively. The vectors $\mathbf{a}_{rx, BS}$ and $\mathbf{a}_{tx, BS}$ correspond to the BS's receive and transmit array responses.

The antenna response vectors $\mathbf{a}_{\mathrm{r}}(\theta^{(\text{rx})}) \in \mathbb{C}^{N_r \times 1}$ and $\mathbf{a}_{\mathrm{t}}(\theta^{(\text{tx})}) \in \mathbb{C}^{N_t \times 1}$ represent the receive and transmit array responses, respectively, for signals arriving from angle $\theta^{(\text{rx})}$ and departing toward angle $\theta^{(\text{tx})}$. For a uniform linear array (ULA), these can be expressed as:
\begin{equation} 
	[\mathbf{a}(\theta)]_n = e^{j\frac{2\pi}{\lambda}(n-1)d\sin(\theta)}, \quad n = 1, 2, \ldots, N,
	\label{eq:antenna_response}
\end{equation}
where $\lambda$ is the wavelength, $d$ is the antenna spacing, and $N$ is the number of antennas.

The Fisher Information Matrix (FIM) $\mathbf{J}(\boldsymbol{\xi}) \in \mathbb{R}^{M \times M}$ characterizes the sensing performance, where $\boldsymbol{\xi} = [\xi_1, \ldots, \xi_M]^T$ contains the $M$ unknown parameters to be estimated (e.g., ranges, velocities, angles). The $(m,n)$-th element of the FIM is defined as:
\begin{equation} \label{FIMStructure}
	[\mathbf{J}(\boldsymbol{\xi})]_{m,n} = \mathbb{E}\left[\frac{\partial \ln p(\mathbf{y}|\boldsymbol{\xi})}{\partial \xi_m} \frac{\partial \ln p(\mathbf{y}|\boldsymbol{\xi})}{\partial \xi_n}\right],
\end{equation}
where $p(\mathbf{y}|\boldsymbol{\xi})$ is the likelihood function of the received signal $\mathbf{y}$ given the parameter vector $\boldsymbol{\xi}$. The CRB provides the lower bound on estimation error variance as $\text{Var}(\hat{\xi}_m) \geq [\mathbf{J}^{-1}(\boldsymbol{\xi})]_{m,m}$ for any unbiased estimator $\hat{\xi}_m$ of parameter $\xi_m$.

The ISAC receiver must simultaneously decode user data from the communication term and estimate the target parameters $\{\alpha_l, \tau_l, f_{D,l}, \phi_l\}_{l=1}^L$ for environment awareness. This dual functionality creates a fundamental trade-off between communication performance and sensing accuracy.

A common optimization formulation for ISAC system design balances these competing objectives as:
\begin{equation}
	\max_{\mathbf{W}}\ R(\mathbf{W})\quad\mathrm{s.t.}\;
	\mathrm{CRB}_{\text{sensing}}(\mathbf{W})\le\varepsilon,\;
	\|\mathbf{W}\|_F^{2}\le P,
	\label{eq:wsum}
\end{equation}
where $\mathbf{W} \in \mathbb{C}^{N_t \times K}$ is the precoding matrix for $K$ communication users, $R(\mathbf{W}) = \sum_{k=1}^K \log_2(1 + \text{SINR}_k(\mathbf{W}))$ represents the achievable sum-rate with $\text{SINR}_k(\mathbf{W})$ being the signal-to-interference-plus-noise ratio for user $k$, $\varepsilon > 0$ is a sensing accuracy threshold, and $P$ is the total transmit power budget.

The sensing performance constraint is characterized by the CRB, which provides a lower bound on the estimation error variance:
\begin{equation}
	\mathrm{CRB}_{\text{sensing}}(\mathbf{W}) = \text{tr}(\mathbf{J}^{-1}(\mathbf{W})),
	\label{eq:crbb}
\end{equation}
where $\mathbf{J}(\mathbf{W})$ is the FIM for the target parameters, and $\text{tr}(\cdot)$ denotes the matrix trace operation. The FIM quantifies the amount of information the received signal carries about the unknown parameters.

This optimization framework illustrates the core ISAC design challenge: maximizing communication throughput while maintaining sufficient sensing accuracy within power constraints. The solution involves careful waveform design and resource allocation that can exploit the \emph{coordination gain} between sensing and communication functionalities, where sensing information can improve channel estimation and beamforming, while communication signals provide additional degrees of freedom for target parameter estimation.

Beyond serving as a generic constraint, the FIM structure in Eq.~\eqref{FIMStructure} provides concrete design insight by revealing how different aspects of the communication signal impact estimation of specific sensing parameters. In the joint ISAC model of Eqs. \eqref{eq:comm_model}, \eqref{eq:sensing_model}, and \eqref{eq:antenna_response}, the entries of the FIM can be expressed in terms of sensitivities of the received signal with respect to the unknown parameters (e.g., range, velocity, angle). For \emph{range} (or time-delay) estimation, the dominant terms in the FIM scale with the effective occupied bandwidth and the temporal diversity of the signal, this is made explicit for CP-OFDM in Eq. \eqref{eq:crb}, where the CRB for delay decreases with the squared effective bandwidth $B_{\mathrm{eff}}^2$ and with the cubic factor $(N-1)N(N+1)$ capturing the number of useful samples per symbol. Thus, increasing bandwidth, using more active subcarriers, or extending the observation length directly strengthens the corresponding FIM entries and tightens the range CRB, at the expense of higher peak-to-average power ratio (PAPR) and tighter hardware requirements. For \emph{velocity} (Doppler) estimation, the FIM is driven by how often and over what interval the phase of the signal is observed: longer coherent processing intervals, denser pilot placement across OFDM symbols, and carefully chosen pulse repetition intervals all increase sensitivity to Doppler-induced phase rotations and reduce the CRB on $f_D$, but they also increase latency and memory demands. For \emph{angle} estimation, the relevant FIM entries depend on the spatial structure of the array-response vectors $a_{\mathrm{tx}}(\cdot)$ and $a_{\mathrm{rx}}(\cdot)$: larger apertures, more antennas, and richer angular diversity (e.g., via multi-beam illumination or multi-RIS paths) yield steeper derivatives with respect to angle and therefore tighter angular CRBs, at the cost of additional RF chains and calibration complexity.

To adapt these fundamental theoretical models into practical 6G deployments, it is necessary to account for the physical limitations and dynamic conditions inherent in operational environments. While the signal models in Eq.~\eqref{eq:comm_model} and Eq.~\eqref{eq:sensing_model} provide a fundamental theoretical basis, several implicit assumptions limit their direct applicability to practical deployments. First, the models assume perfect knowledge of the number of propagation paths $P$ and targets $L$, which in practice must be estimated jointly with the channel parameters, introducing model order uncertainty~\cite{2.1.3, 2.1.4}. Second, the path gains $\beta_p$ and $\alpha_l$ are modeled as deterministic scalars, neglecting small-scale fading variations due to multipath scattering, which in realistic channels exhibit Rayleigh or Rician distributions~\cite{2.1.5, 2.1.6}. Third, perfect synchronization between transmitter and receiver is assumed, whereas bistatic and distributed ISAC systems suffer from timing offsets, carrier frequency offsets, and phase noise that introduce additional unknown parameters into the observation model~\cite{2.1.7}. Fourth, the additive noise terms $\mathbf{n}_c(t)$ and $\mathbf{n}_s(t)$ are assumed white Gaussian, which may not hold in the presence of colored interference from co-channel users or non-Gaussian clutter returns from extended targets~\cite{2.1.8, 4.7.3.2}.

The CRB derived in Eq.~\eqref{eq:crbb} provides a lower bound on estimation error variance under four critical conditions: i) the parameters $\boldsymbol{\xi}$ are deterministic and time-invariant during the observation interval, ii) the estimator is \textit{unbiased}, iii) the likelihood function $p(\mathbf{y}|\boldsymbol{\xi})$ is differentiable with respect to $\boldsymbol{\xi}$~\cite{2.1.9}, and iv) the system operates in the high signal-to-noise ratio (SNR) regime where the bound is asymptotically tight. However, these conditions are frequently violated in high-mobility 6G scenarios. In vehicular networks operating at millimeter-wave frequencies, targets move at velocities exceeding 100~km/h, causing Doppler shifts and ranges to vary by several wavelengths within a single OFDM symbol duration~\cite{2.1.10}. Similarly, in UAV-assisted ISAC and RIS-enhanced systems, both the sensing platform and targets exhibit rapid three-dimensional motion, rendering the bistatic geometry time-variant and violating assumption~(i)~\cite{2.1.11}. Under such conditions, the classical CRB underestimates the true estimation error because it does not account for the \textit{mismatch} between the assumed static model and the actual dynamic evolution of $\boldsymbol{\xi}$~\cite{2.1.9, 2.1.12}.

To address these limitations, two categories of extensions are necessary. For the signal model, practical systems require \textit{extended observation models} that explicitly incorporate synchronization errors, time-varying channels, and model order uncertainty. For instance, in bistatic ISAC with clock asynchronism, the sensing model in Eq.~\eqref{eq:sensing_model} must be augmented with unknown timing offset $\tau_{\text{TO}}$ and carrier frequency offset $f_{\text{CFO}}$, yielding a modified received signal:
	\begin{equation}
		\begin{split}
			\mathbf{y}_s(t) = & \sum_{l=1}^{L} \alpha_l e^{j2\pi (f_{D,l} + f_{\text{CFO}}) t} \mathbf{a}_{rx, BS}(\phi_{rx, l}) \\
			& \times \mathbf{a}_{tx, BS}^T(\phi_{tx, l}) \mathbf{x}(t - \tau_{s,l} - \tau_{\text{TO}}) + \mathbf{n}_s(t),
		\end{split}
		\label{eq:modified_sensing_model}
	\end{equation}
which introduces additional parameters that must be jointly estimated or calibrated~\cite{2.1.7}. Beyond synchronization errors, time-varying channels due to high mobility require \textit{state-space formulations} to capture parameter evolution. In such cases, the static parameter model can be replaced with a dynamic state-space representation: $\boldsymbol{\xi}_k = \mathbf{f}(\boldsymbol{\xi}_{k-1}, \mathbf{w}_k)$ and $\mathbf{y}_k = \mathbf{h}(\boldsymbol{\xi}_k, \mathbf{v}_k)$, where $\mathbf{f}(\cdot)$ represents the target motion dynamics (e.g., constant velocity or constant acceleration) and $\mathbf{w}_k, \mathbf{v}_k$ are process and measurement noise~\cite{2.1.14}. For performance bounds, the PCRB provides a Bayesian alternative to the classical CRB by treating $\boldsymbol{\xi}$ as a random variable with prior distribution $p(\boldsymbol{\xi}_0)$ and evolution model $p(\boldsymbol{\xi}_k | \boldsymbol{\xi}_{k-1})$, yielding a recursive bound suitable for sequential estimation in dynamic environments: $\mathbb{E}[\|\hat{\boldsymbol{\xi}}_k - \boldsymbol{\xi}_k\|^2 | \mathbf{y}_{1:k}] \geq \text{tr}(\mathbf{J}_k^{-1})$, where $\mathbf{J}_k$ is the Bayesian information matrix~\cite{2.1.13}. These extended formulations align theoretical analysis with the operational realities of 6G ISAC systems characterized by high mobility, asynchronous operation, and uncertain propagation environments~\cite{4.3.1}.

Finally, regarding the high-SNR condition, it is crucial to recognize that the CRB is a local bound that predicts the performance of efficient estimators only in the asymptotic region. In practical ISAC systems operating under power constraints or severe path loss (low-to-medium SNR), the so-called \textquotedblleft threshold effect\textquotedblright $\,$occurs. In this regime, the mean squared error (MSE) of realizable estimators (e.g., Maximum Likelihood) deviates sharply from the CRB due to the presence of large outlier errors \cite{2.1.15}. Consequently, the CRB may provide an overly optimistic benchmark in these scenarios. To rigorously characterize performance limits across all SNR regimes, global bounds such as the Barankin Bound or the Ziv-Zakai Bound are required, as they capture the transition from the noise-dominated to the signal-dominated region \cite{2.1.16}.

\subsection{Joint Design and Coordination Approaches}

ISAC architectures fall into two broad classes.
\subsubsection{Joint (Co‑Design) ISAC}
Communication and sensing share the same waveform and hardware, requiring joint optimization of waveforms, beamforming, and resource allocation. Recent research and standardization efforts emphasize that embedding pilot symbols within data blocks can significantly enhance both channel estimation accuracy and target detection performance, effectively leveraging the dual use of transmitted signals \cite{2.2.1}. Moreover, advanced waveform designs such as dual-domain superposition of delay-Doppler signals on cyclic-prefix (CP) OFDM have demonstrated remarkable improvements in sensing precision, reducing the CRB for range estimation by up to $20$ dB without compromising communication throughput \cite{2.2.2}. These innovations highlight the potential of joint ISAC systems to deliver superior spectral efficiency and sensing capability, positioning them as a cornerstone technology for future 6G networks where seamless coexistence and mutual enhancement of sensing and communication are paramount.
\subsubsection{Coordination‐Based ISAC} 
A more modular approach is adopted by employing distinct signals for sensing and communication functions, with coordinated use of time, frequency, or spatial resources. This method partitions the cell sector into dedicated sensing and communication beams, dynamically allocating power to ensure user rate guarantees and simultaneously controlling clutter and interference in the sensing domain. Such coordination schemes offer the advantage of backward compatibility with existing communication standards, facilitating incremental deployment and integration into current network infrastructures. However, this separation inherently sacrifices some spectral efficiency compared to joint designs, as resources are partitioned rather than fully shared \cite{2.2.3}. Despite this trade-off, coordination-based ISAC remains a practical and effective solution, especially in scenarios where legacy compatibility and system simplicity are prioritized. The dynamic resource management and beam partitioning techniques underpinning coordination-based ISAC are actively being explored to balance sensing accuracy and communication quality of service, thereby enabling robust and flexible network operation in complex environments \cite{2.2.4}.

\subsection{Monostatic, Bistatic, and Multistatic Architectures}

\subsubsection{Monostatic ISAC} 
Systems that integrate the transmitter and receiver within the same physical location greatly simplify synchronization and clock alignment since both sensing and communication share a common timing reference. However, this co-location introduces the challenge of strong self-interference, where the transmitted signal can leak into the receiver and mask the weak echoes reflected from targets. To overcome this, advanced self-interference cancellation techniques are necessary. Recent developments in full-duplex mmWave prototypes have demonstrated the ability to achieve over $40$ dB of analog cancellation, enabling simultaneous high-throughput communication links at gigabit-per-second speeds while maintaining effective sensing capabilities. This balance makes monostatic ISAC particularly suitable for applications requiring compact hardware and low latency, such as vehicular radar and base station sensing \cite{2.3.1.1, 2.3.1.2, 4.7.3.2}.
\subsubsection{Bistatic ISAC} 
When the transmitter and receiver are placed separately on distinct nodes or sites, this spatial separation naturally addresses the self-interference issues found in monostatic setups. The sensing range in such systems follows the bistatic range equation:
    \begin{equation}
    R_{\mathrm b}=d_{\mathrm{Tx,t}}+d_{\mathrm{t,Rx}}-d_{\mathrm{Tx,Rx}},
    \label{eq:bistatic}
    \end{equation}
where $d_{\mathrm{Tx,t}}$ and $d_{\mathrm{t,Rx}}$ represent the distances from the transmitter to the target and from the target to the receiver, respectively, and $d_{\mathrm{Tx,Rx}}$ is the baseline distance between transmitter and receiver. While this geometry enhances sensing coverage and flexibility, bistatic ISAC requires extremely precise synchronization between the spatially separated nodes, often at nanosecond-level accuracy, to correctly interpret the timing of received echoes and maintain coherent processing. Such stringent synchronization demands increase system complexity but enable distributed sensing applications like UAV networks and cooperative smart city monitoring \cite{2.3.2.1, 2.3.2.2, 2.3.2.3}.
\subsubsection{Multistatic ISAC} 
Leveraging the extension of the bistatic model, deploying numerous transmitters and receivers in a distributed manner capitalizes on spatial diversity for better detection performance and reduced fading. Each transmitter-receiver pair provides an independent measurement of the target scene, and the fusion of these measurements enhances overall sensing accuracy and robustness. However, the increased number of RF chains and hardware components raises concerns about power consumption and system complexity. Techniques such as RF-chain selection have been shown to effectively reduce power usage; for instance, a $128$-antenna multistatic network can achieve approximately $35$\% power savings while still meeting desired detection probability targets. This energy-efficient operation is vital for scalable and dense $6$G deployments where large-scale environmental monitoring is necessary without incurring prohibitive energy costs \cite{2.3.3.1, 2.3.3.2, 2.3.3.3}.

A concise comparison of these ISAC architecture types, including their configurations, advantages, disadvantages, and typical use cases, is summarized in Table~\ref{tab:isac_system_types}.

\begin{table*}[!ht]
\centering
\caption{Comparison of ISAC Architecture Types}
\label{tab:isac_system_types}
\renewcommand{\arraystretch}{1.3}
\setlength{\tabcolsep}{4pt}
\begin{tabular}{|>{\bfseries}l|p{3.8cm}|p{3.8cm}|p{3.8cm}|p{3.8cm}|}
\hline
\rowcolor[HTML]{E6E6E6}
System Type & \textbf{Configuration} & \textbf{Advantages} & \textbf{Limitations} & \textbf{Typical Use Cases} \\
\hline
\hline
\makecell{Monostatic\\\cite{2.3.1.1, 2.3.1.2, 4.7.3.2}} & Transmitter and receiver are co-located (e.g., at the base station). & Simplified synchronization and processing pipeline. & Suffers from self-interference; limited spatial diversity. & Automotive radar, base station environmental sensing. \\
\hline
\makecell{Bistatic\\\cite{2.3.2.1, 2.3.2.2, 2.3.2.3}} & Transmitter and receiver are spatially separated. & Reduces self-interference; improves spatial diversity and flexibility. & Requires accurate synchronization and timing control between nodes. & UAV sensing, distributed surveillance networks. \\
\hline
\makecell{Multistatic\\\cite{2.3.3.1, 2.3.3.2, 2.3.3.3}} & Multiple distributed transmitters and receivers work cooperatively. & Enables high-resolution sensing via spatial diversity and redundancy. & High system complexity; requires tight coordination and data fusion. & Large-scale environment monitoring, smart infrastructure. \\
\hline
\end{tabular}
\end{table*}

\subsection{ISAC Waveform Classification}
Waveform design is critical in ISAC to simultaneously support high data rates and accurate sensing. Table~\ref{tab:isac_waveforms} summarizes the main classifications of ISAC waveforms.

For CP-OFDM with fast fourier transform (FFT) size $N$ and effective bandwidth $B_{\text{eff}} = (N_{\text{active}}-1)\Delta f$, where $N_{\text{active}}$ represents the number of active subcarriers and $\Delta f$ is the subcarrier spacing, the range estimation Cramér-Rao bound \cite{2.4.1} is
\begin{equation}
\mathrm{CRB}_{\tau}=\frac{6}{(2\pi B_{\text{eff}})^2 \mathrm{SNR}\, (N-1)N(N+1)}.
\label{eq:crb}
\end{equation}
This bound represents the theoretical lower limit on time delay estimation variance, where the cubic polynomial term $(N-1)N(N+1)$ reflects the degrees of freedom available in the OFDM symbol structure for parameter estimation. The effective bandwidth $B_{\text{eff}}$ also determines the theoretical range resolution $\Delta r = \frac{c}{2B_{\text{eff}}}$, which characterizes the system's ability to distinguish between closely spaced targets.

The sensing performance of ISAC waveforms is fundamentally limited by their autocorrelation properties. For CP-OFDM systems, the periodic autocorrelation function (P-ACF) is defined as:
\begin{equation}
R_{xx}^{(P)}(\tau) = \frac{1}{T} \int_0^T x(t)x^{*}(t-\tau) dt,
\label{eq:pacf}
\end{equation}
where $T$ is the OFDM symbol period including cyclic prefix. Lower sidelobe energy in $R_{xx}^{(P)}(\tau)$ directly improves range ambiguity performance by reducing false target detections. The integrated sidelobe level (ISL) quantifies this effect as $\text{ISL} = \sum_{\tau \neq 0} |R_{xx}^{(P)}(\tau)|^2$, where lower ISL values indicate superior range disambiguation capability.

Recent theoretical analysis in \cite{2.4.2} rigorously proves that CP-OFDM achieves the globally optimal performance by minimizing the expected integrated sidelobe level among all communication-centric ISAC waveforms using quadrature amplitude modulation (QAM)/phase-shift keying (PSK) constellations. This optimality result establishes CP-OFDM as the preferred waveform for ISAC applications requiring both high-rate communication and accurate ranging.

\begin{table*}[!ht]
\centering
\caption{Classification of ISAC Waveforms and Their Characteristics}
\label{tab:isac_waveforms}
\renewcommand{\arraystretch}{1.35}
\setlength{\tabcolsep}{6pt}
\begin{tabular}{|>{\bfseries}p{3.5cm}|p{4.3cm}|p{4.4cm}|p{4.4cm}|}
\hline
\rowcolor[HTML]{E6E6E6}
Waveform Type & \textbf{Description} & \textbf{Sensing Characteristics} & \textbf{Communication Characteristics} \\
\hline
\hline
CP-OFDM \cite{T31} & Standard OFDM with cyclic prefix, widely used in 4G/5G. & Low range sidelobes, good delay resolution, FFT-compatible for range-Doppler processing. & High spectral efficiency, robust to multipath fading, easy integration into existing networks. \\
\hline

Code-division OFDM (CD-OFDM) \cite{T32} & Applies code-division multiple access (CDMA) spreading to OFDM subcarriers. & Improved sensing resolution via code diversity, enhanced target separability. & Supports multi-user access, robust to narrowband interference, improved security. \\
\hline

Single-carrier with CP (SC-CP) \cite{T33} & Single-carrier waveform with cyclic prefix to mitigate ISI. & Low Doppler sidelobes, ideal for high-speed target tracking. & Lower PAPR than OFDM, well-suited for power-limited uplink transmission. \\
\hline

OTFS \cite{T34} & Modulates in delay–Doppler domain, resilient to fast-varying channels. & High Doppler resilience, sharp ambiguity function profile. & Maintains robustness in mobility-rich environments, optimal for vehicular use. \\
\hline

frequency-modulated continuous wave (FMCW)/Chirp–OFDM Hybrid \cite{T35} & Integrates FMCW chirps with OFDM for dual functionality. & Offers high range resolution via chirp modulation, excellent for fine-grained mapping. & Preserves OFDM compatibility, maintains robust data transmission. \\
\hline

Dual-domain ISAC Waveforms \cite{T36} & Overlays sensing signals in delay–Doppler over OFDM base. & Up to 10–20 dB CRB improvement, facilitates integrated radar-comm designs. & Preserves throughput, allows flexible trade-off tuning. \\
\hline

Spread spectrum ISAC (e.g., DSSS) \cite{T37} & Wideband waveform using direct-sequence spreading. & Enhanced detection under low-SNR, increased processing gain. & High resistance to interference, suitable for covert and low power wide area network (LPWAN) communication. \\
\hline

Frequency-hopping ISAC \cite{T38} & Rapid frequency switching across bands for resilience. & Anti-jamming and low probability of intercept sensing. & Robust in contested environments, improved spectral agility. \\
\hline

Adaptive waveforms \cite{T39} & Reconfigurable waveforms responsive to context/environment. & Adaptive sensing precision and latency based on scenario. & Energy-efficient and quality of service (QoS)-aware, spectrum-adaptive modulation. \\
\hline

\end{tabular}
\end{table*}

\section{Key Enabling Technologies} \label{KET}
This section details the primary technologies that enable ISAC in 6G networks, focusing on antenna systems, waveform design, sensing-aware resource allocation, and machine learning/AI techniques.

\subsection{Antenna Technologies} \label{AT}
\subsubsection{Massive Multiple-Input Multiple-Output}

\begin{figure}[t]
	\includegraphics[width=8cm,height=0.25\textheight]{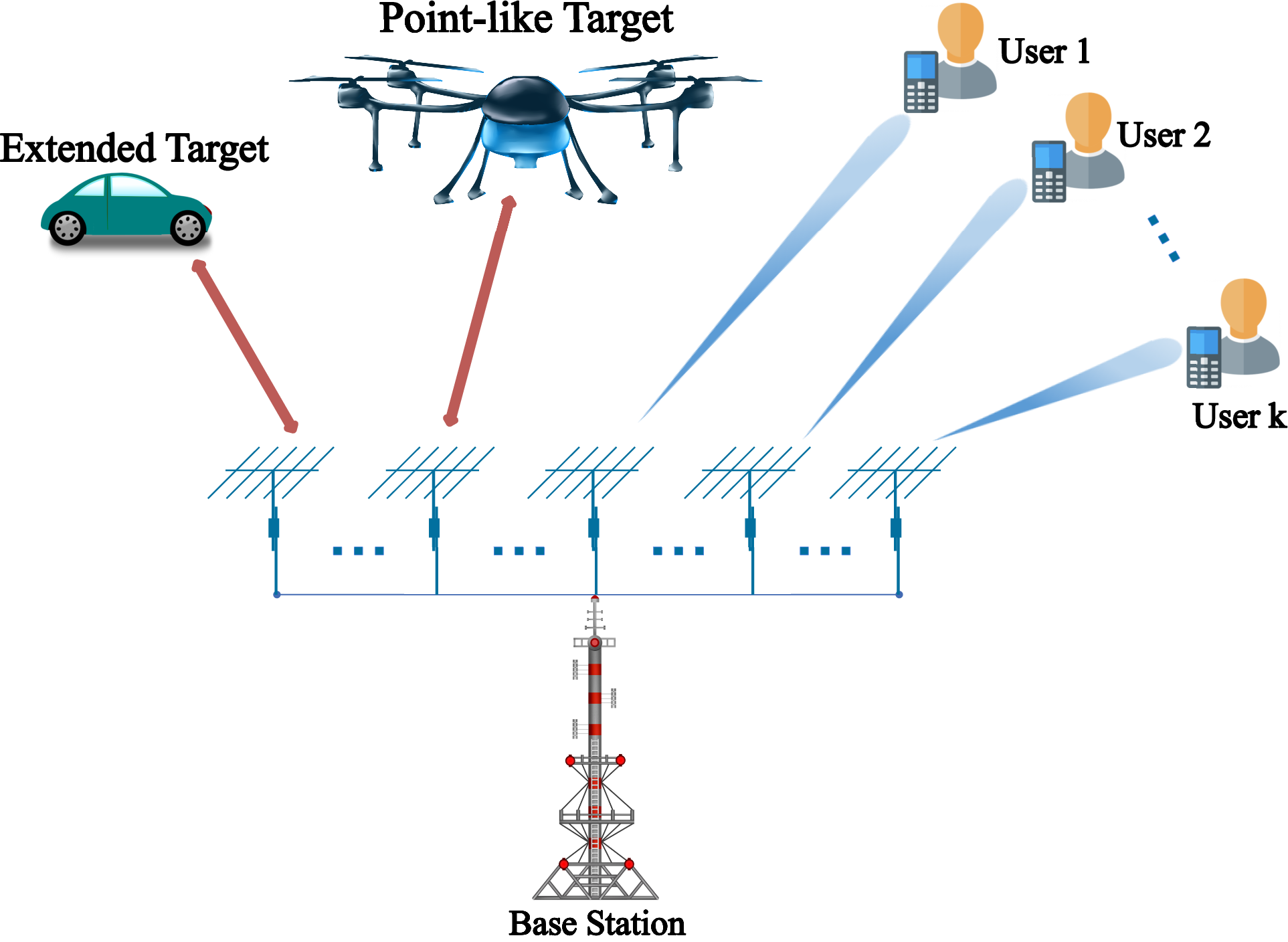}
	\centering
	\caption{Illustration of Massive MIMO-ISAC system.}
	\label{fig:MIMO-ISAC}
\end{figure}

Massive MIMO is a foundational technology for ISAC in 6G systems. It involves deploying a large number of antennas, often tens to hundreds, at the base station to enable high spatial resolution and directional beamforming. This large array configuration allows the system to serve multiple users simultaneously while also probing the environment for sensing, using the same hardware and waveform as shown in Fig.~\ref{fig:MIMO-ISAC}. The ability to form narrow beams improves SNR, enhances user throughput, and enables accurate target detection, range estimation, and angular localization. This capability is particularly valuable in use cases such as autonomous vehicles, aerial drones, and smart infrastructure, where real-time environmental awareness and high-capacity connectivity must coexist \cite{3.1.1.1},\cite{3.1.1.2}.

The received signal at a base station equipped with massive MIMO for ISAC sensing applications follows the sensing signal model derived in Section~\ref{subsec:DBP}, specifically Eq.~\eqref{eq:sensing_model}. In this model, the massive antenna array gain is captured in the high-dimensional steering vectors $\mathbf{a}_{rx, BS}$ and $\mathbf{a}_{tx, BS}$. The large number of antennas results in asymptotically orthogonal channel vectors, which allows the BS to distinguish the sensing echoes from user uplink signals and interference with high precision. Simultaneously, the downlink communication performance is governed by Eq.~\eqref{eq:comm_model}, where the massive array gain maximizes the received signal power at the UE via beamforming. This dual capability reflects how massive MIMO enables the joint decoding of user data and estimation of target parameters (range $\tau_{s,l}$, velocity $f_{D,l}$, and angle $\phi_{l}$) within the same infrastructure \cite{3.1.1.3}.

In practical terms, massive MIMO systems employ beamforming techniques such as zero-forcing or minimum mean square error (MMSE) to spatially separate users and targets, improving accuracy and reliability. This capability makes massive MIMO a critical enabler of ISAC, providing the necessary spatial diversity and precision for next-generation intelligent and perceptive networks \cite{3.1.1.4}.

\subsubsection{Reconfigurable Intelligent Surfaces}
RIS, also known as intelligent reflecting surfaces (IRS), are now an essential technology for next-generation wireless networks. RIS are meta-surfaces designed to change the amplitude, phase, and polarization of incident electromagnetic waves. They consist of a wide range of low-cost, programmable components. It offers unparalleled flexibility in forming radio channels for communication and sensing applications by dynamically modifying these properties, intelligently controlling signal propagation in the wireless environment \cite{3.1.2.1}.

One of the most appealing aspects of RIS is its ability to reconfigure the wireless environment in real time. It can actively build or improve NLOS links by reflecting and focusing signals toward specific locations, as opposed to conventional infrastructure, which gradually adjusts to environmental changes.  This feature is particularly useful in indoor or urban settings where obstacles frequently obstruct direct line-of-sight routes. One of the primary issues in dense and complex transmission environments is addressed by RIS, which significantly improves coverage and reliability for data transmission and environmental sensing by enabling NLOS connectivity \cite{3.1.2.2}.

\begin{figure}[t]
	\includegraphics[width=\linewidth,height=0.2\textheight]{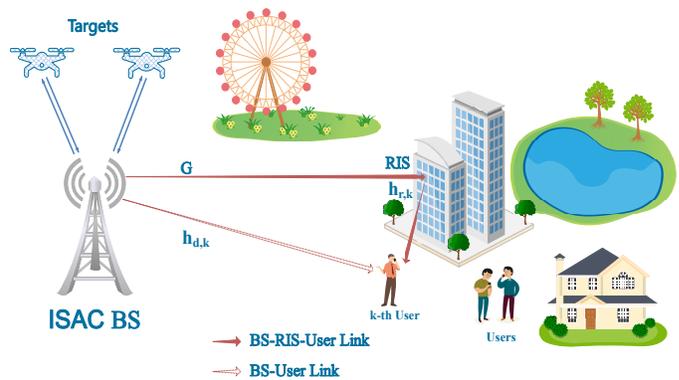}
	\centering
	\caption{RIS-assisted ISAC system}
	\label{fig:RIS-ISAC}
\end{figure}

RIS is also necessary for joint beamforming. It can shape sensing beams for environmental surveillance while also routing communication signals to their intended users by precisely controlling the reflected wavefronts, as illustrated in Fig.~\ref{fig:RIS-ISAC}. The system operates through multiple channel paths including the direct channel coefficient $h_{d,k}$ from the base station to the k-th user, the RIS channel coefficient $G$ between the ISAC base station and RIS, and the reflected channel coefficient $h_{r,k}$ from RIS to users, enabling both direct and reflected communication paths for enhanced coverage. This dual functionality not only improves energy and spectral efficiency but also allows for sophisticated applications such as object tracking, gesture recognition, and high-precision localization.  It is ideal for meeting the real-time needs of ISAC systems due to its programmability, which allows for quick adaptation to changing circumstances \cite{BEAMRIS}.

Recent research has shown that combining RIS with cell-free massive MIMO architectures has several advantages \cite{3.1.2.4, 3.1.2.5, 3.1.2.6, 3.1.2.7}. Large-scale antenna arrays and dispersed RIS panels collaborate in these deployments to create a highly reconfigurable and spatially diverse wireless infrastructure.  This integration has resulted in significant gains in spectral efficiency, energy consumption, and sensing accuracy, particularly in situations involving extreme multipath fading or shadowing.  Networks can use RIS's spatial diversity and programmable control to achieve finer-grained beam steering, environmental mapping, and interference management, all of which are required for ISAC to function reliably in 6G and beyond.

Novel forms of RIS, like simultaneously transmitting and reflecting RIS (STAR-RIS) and active RIS, make ISAC even better by allowing for simultaneous full-duplex operations, more accurate sensing coverage, and energy transfer \cite{3.1.2.8}. Beyond traditional MIMO and RIS, Table~\ref{tab:advanced_antennas} lists additional antenna technologies, highlighting their distinct roles and capabilities in ISAC and next-generation wireless systems.

Despite the individual promise of these advanced antenna architectures, their heterogeneous integration into a unified multi-vendor 6G network introduces significant interoperability challenges. First, \textit{codebook incompatibility} arises between holographic and conventional phased arrays, holographic surfaces rely on continuous aperture impedance modulation which renders standard DFT-based precoding codebooks (designed for discrete element spacing) ineffective, potentially leading to severe beam misalignment in blind multi-vendor pairing \cite{3.1.2.8a}. Second, \textit{beam squint effects} in ultra-wideband Terahertz (THz) arrays create frequency-dependent pointing errors. Without standardized wideband beam-tracking protocols, a THz transmitter from one vendor may fail to maintain lock with a lens-antenna receiver from another during frequency agile sensing operations \cite{3.1.2.8b}. Third, \textit{near-field mismatch} becomes critical when integrating large-aperture lens or holographic arrays with legacy far-field equipment, the wavefront curvature assumptions differ, requiring unified CSI reporting formats that can adaptively switch between near-field and far-field models to prevent degradation in sensing resolution \cite{3.1.2.8c}. Addressing these compatibility issues requires the development of technology-agnostic interfaces and capability exchange signaling within future 3GPP and O-RAN specifications.

\begin{table*}[!ht]
\centering
\caption{Other Advanced Antenna Technologies for 6G and ISAC}
\label{tab:advanced_antennas}
\renewcommand{\arraystretch}{1.35}
\setlength{\tabcolsep}{8pt}
\begin{tabular}{|>{\bfseries}p{2.5cm}|p{6.5cm}|p{7.2cm}|}
\hline
\rowcolor[HTML]{E6E6E6}
Antenna Type & \textbf{Key Features} & \textbf{Benefits and Applications} \\
\hline
\hline
\makecell{Holographic\\\cite{1.0, 3.1.2.9}}& 
Ultra-dense metasurface arrays, either continuous or discrete, with software-defined beamforming and dynamic wavefront shaping capabilities. & 
Enables fine-grained spatial control, adaptive beam steering, and high-resolution environmental sensing for ISAC and 6G. \\
\hline

\makecell{Lens\\\cite{2.2.2, 3.1.2.9}} & 
Utilizes electromagnetic or dielectric lenses, often in conjunction with phased arrays, to focus and separate signal paths efficiently. & 
Delivers high directivity and low interference, making it ideal for spatial multiplexing at mmWave and THz frequencies. \\
\hline

\makecell{Terahertz\\\cite{3.1.2.10, 3.1.2.11}}& 
Operates in the 0.1–10 THz range using advanced plasmonic, photonic, or nano-antenna designs for extreme miniaturization. & 
Supports ultra-high throughput, sub-millimeter resolution for sensing, and novel applications in 6G environments. \\
\hline

\end{tabular}
\end{table*}

\subsection{Waveform Design}
Waveform design is a fundamental challenge in ISAC, as it demands a trade-off between two conflicting objectives. On one hand, communication relies on spectral efficiency and orthogonality to maximize data rates. On the other hand, sensing requires waveforms with specific ambiguity function properties to ensure high accuracy in range and velocity estimation. Bridging these requirements typically involves three design strategies: sensing-centric, communication-centric, and joint waveform optimization.

While academia has explored a wide variety of novel waveforms across these categories, standardization bodies have prioritized backward compatibility and stability for the initial 6G specifications.
\subsubsection{Orthogonal Frequency Division Multiplexing (OFDM)} \label{OFDM}
OFDM remains the dominant waveform for ISAC systems. Crucially, the 3GPP RAN1 working group has officially agreed that CP-OFDM and DFT-s-OFDM will serve as the baseline waveforms for the 6G air interface \cite{3gpp_waveform_agreement}. This decision prioritizes the proven robustness and spectral efficiency of OFDM for both uplink and downlink channels, ensuring smooth evolution from 5G Advanced to 6G. Consequently, the primary implementation of ISAC in early 6G networks will rely on sensing algorithms compatible with the standardized CP-OFDM grid.

Other advanced ISAC waveforms covered in this tutorial, such as OTFS, FBMC, and AFDM, are therefore positioned not as immediate standardization targets, but as advanced research candidates or potential ``Next-Evolution Waveforms'' (NEW) for specific high-mobility niches or long-term evolution beyond the initial 6G releases.

Integrating pulse-Doppler radar principles into the OFDM resource grid allows for high-resolution sensing, which is one of OFDM's main advantages in ISAC. By performing matched filtering on the received pulses, the sensing function extracts correlation peaks to estimate target parameters. Specifically, the wide bandwidth of the OFDM signal improves range resolution, while the phase differences between successively received pulses contribute to velocity computation and accurate Doppler shift measurement. Radar detection and communication data transmission can occur simultaneously due to the dual utilization of OFDM subcarriers, which is an essential synergy for applications like smart environments and vehicle networks where both tasks must successfully coexist \cite{3.2.1.3, 3.2.1.4}.

Nonetheless, OFDM-based ISAC encounters substantial challenges, notably a trade-off between sensing precision and communication throughput. This trade-off is governed by the waveform's ambiguity function, which determines the resolution and precision of delay and Doppler estimates. Enhancing this trade-off necessitates advanced resource allocation techniques, including adaptive power allocation and subcarrier selection, to reduce the Cramér-Rao Bound (CRB) for delay and Doppler estimates. Minimizing the CRB improves sensing performance by decreasing estimation errors while maintaining communication quality. Proposed solutions to these challenges include advanced waveform designs like Discrete Fourier Transform (DFT)-spread OFDM integrated with Index Modulation (IM). These designs enhance sensing performance by manipulating the ambiguity function to attain locally optimal auto-correlation characteristics in specified areas of the delay-Doppler domain, while ensuring reliable communication throughput \cite{3.2.1.4, 3.2.1.5, 3.2.1.6}.

One of the innovations in OFDM-based ISAC is Superposed Index Modulated OFDM (S-IM-OFDM), which increases sensing capabilities without increasing power usage. By integrating sensing-oriented signals into the OFDM waveform using energy-efficient index modulation, S-IM-OFDM improves detection performance, especially in dynamic, time-varying channels. Furthermore, this technique enables Doppler compensation using measurable parameters, preserving communication reliability in demanding circumstances \cite{3.2.1.7}. In addition, Directional Modulation (DM) techniques combined with OFDM (OFDM-DM) have been proposed to reduce interference from non-target directions and enhance security by blocking eavesdropping via unanticipated paths \cite{3.2.1.8}. These developments demonstrate how OFDM waveforms are optimized for simultaneous sensing and communication capabilities while maintaining a balance between performance, power efficiency, and complexity.

\subsubsection{Frequency Modulated Continuous Wave and Hybrid Waveforms}
FMCW radar has long been foundational in automotive sensing due to its precise range and velocity estimation at relatively low hardware cost and complexity. Its hallmark is the linear modulation of frequency over time, which enables the robust estimation of target distance and motion by measuring the time delay and frequency shift between transmitted and received signals. As vehicle and sensor networks \cite{SDORP} increasingly demand joint communication and sensing \cite{Poised, ISAC-A, ISAC-F}, FMCW technology is being actively adapted for ISAC scenarios. This adaptation leverages the dual potential of the FMCW signal: it retains high range and velocity accuracy for sensing while enabling simultaneous data embedding for communication, resulting in efficient use of scarce radio frequency resources allocated to automotive and industrial radar systems \cite{3.2.2.1},\cite{3.2.2.2}.

Despite these strengths in sensing, FMCW's relatively narrow bandwidth and lower data-carrying capacity present practical limitations when high-throughput connectivity is needed. The area of research is currently focusing on hybrid waveform designs that integrate FMCW with communication-oriented waveforms, specifically OFDM. These hybrid waveforms attempt to integrate the continuous, high-resolution sensing capabilities of FMCW with the higher spectral efficiency and data rates of OFDM. Hybrid approaches leverage the advantages of both FMCW's coherent chirp processing for range and Doppler estimation and OFDM's multicarrier design for high-capacity data transmission through intelligent multiplexing or superposition of these signals \cite{3.2.2.3, 3.2.2.4, 3.2.2.5}.

In densely populated urban areas with multipath propagation, hybrid FMCW-OFDM waveforms exhibit significant potential. The persistent chirp of FMCW enhances resistance to multipath fading and facilitates coherent integration of signal returns, hence enabling accurate object recognition in complex environments.  Simultaneously, OFDM's adaptable subcarrier distribution and guard intervals reduce intersymbol interference and enhance transmission reliability.  These hybrid methodologies provide advanced signal processing frameworks, including joint delay-Doppler estimation methods, which further improve system efficacy in both sensing and communication sectors \cite{3.2.2.3},\cite{3.2.2.5}.

Recent studies indicate that hybrid waveform designs can effectively equilibrate resource allocation between sensing and communication functions to satisfy the requirements of responsive, real-time applications such as V2X, autonomous navigation, and industrial automation. The studies also indicate that sophisticated modulation formats integrated into FMCW can attain higher data rates without substantially compromising sensing efficacy, depending upon careful optimization of the entire signal processing chain \cite{3.2.2.1}.

\subsubsection{Phase-coded and Frequency-coded Waveforms}
Phase-coded waveforms are a modulation technique in which the phase of a carrier signal is systematically altered in accordance with a predetermined coding sequence. This coding improves range resolution and target detection efficacy in radar and ISAC systems, with notable examples such as Barker and Polyphase codes \cite{3.2.3.1},\cite{2.2.2}. On the other hand, frequency-coded waveforms adjust the instantaneous frequency of the carrier, generating signals such as stepped-frequency waveforms or frequency-hopping. These waveforms improve resistance to interference and assist in ambiguity resolution, rendering them ideal for applications that require a low chance of interception or high spectrum efficiency \cite{3.2.3.2},\cite{3.1.2.9}. Both coding methodologies enable flexible trade-offs among detection performance, resolution, and waveform diversity in contemporary wireless and sensing systems.

\subsubsection{Filter Bank Multicarrier and Other Multicarrier Waveforms}
FBMC is a multicarrier modulation technique that segments the frequency spectrum into small sub-bands utilizing filter banks with carefully designed pulse shaping. In contrast to conventional OFDM, FBMC offers enhanced spectral localization and reduces out-of-band emissions due to the elimination of a cyclic prefix and the implementation of offset quadrature amplitude modulation (OQAM) \cite{3.1.2.10},\cite{3.2.4.1}. FBMC and other multicarrier waveforms, including OFDM and universal filtered multicarrier (UFMC), offer significant spectrum effectiveness and resistance to frequency-selective fading, making them essential for next-generation wireless standards and ISAC systems. The selection of multicarrier waveform impacts synchronization sensitivity, complexity, and appropriateness for diverse application requirements in 5G/6G and IoT networks \cite{3.2.4.2},\cite{3.2.4.3}.

\subsection{When and Why to Use Key ISAC Technologies}

The enabling technologies discussed above offer complementary strengths and are not universally optimal. Their suitability depends strongly on the propagation environment, mobility profile, hardware constraints, and dominant performance objective (e.g., sensing resolution vs. throughput vs. coverage). A brief critical comparison is therefore useful to clarify \emph{when} and \emph{why} particular antenna and waveform choices are preferred for different ISAC use cases.
\begin{itemize}
\item \textit{Massive MIMO vs. RIS (and Their Combination):} Massive MIMO arrays at the base station are most effective when the infrastructure can support many active RF chains and digital beamforming, such as in macro-cell deployments for vehicular networks or dense urban hotspots. In these scenarios, massive MIMO provides high angular resolution and multi-user spatial multiplexing, enabling precise target localization and high-capacity links simultaneously, however, it incurs significant hardware cost, power consumption, and calibration complexity. In contrast, RIS offers a largely passive, low-power means to reshape the radio environment, making it particularly attractive for extending coverage and enhancing sensing in NLOS-dominated or blockage-prone environments (e.g., indoor factories, urban street canyons), where deploying additional active base stations is impractical. The main limitations of RIS are its double-fading effect, reliance on accurate CSI for configuration, and the fact that it cannot generate energy on its own. In practice, hybrid deployments, massive MIMO base stations assisted by strategically placed RIS panels, are most suitable for wide-area ISAC scenarios requiring both high-resolution sensing and coverage enhancement, while purely massive MIMO solutions are more appropriate when infrastructure power and cost budgets are less constrained, and purely RIS-assisted architectures are better suited to low-energy, coverage-extension roles.
\item \textit{Baseline vs. Specialized Waveforms:} Given that CP-OFDM and DFT-s-OFDM have been selected as the baseline air-interface waveforms for initial 6G releases, they are the natural choice for \emph{general-purpose} ISAC in cellular deployments, where backward compatibility, spectral efficiency, and mature hardware support are paramount. CP-OFDM-based ISAC is thus best suited for wide-area mobile broadband scenarios, such as urban macro cells where high-capacity communication and precise localization are both required (e.g., location-aware mobile broadband services), where sensing operates as an overlay on standardized communication grids. In contrast, FMCW and OFDM–FMCW hybrid waveforms are preferred when \emph{sensing performance dominates} and very fine range/velocity resolution is required, such as automotive radar, industrial robotics, and short-range positioning; their limitations lie in lower raw data rates and more specialized hardware. OTFS and other delay Doppler waveforms are particularly suited to high-mobility environments (e.g., high-speed rail, UAV corridors) where Doppler robustness is critical, at the cost of increased receiver complexity and currently limited standardization support. Single-carrier CP waveforms are attractive for uplink ISAC from power-constrained user equipment (e.g., IoT sensors, vehicular terminals), where low PAPR is more important than maximum spectral efficiency. Finally, spectrally well-localized multicarrier schemes such as FBMC/UFMC are beneficial in fragmented spectrum or coexistence settings (e.g., ISAC sharing bands with legacy systems), but incur higher implementation complexity and are less aligned with current 3GPP baselines.
\item \textit{Use-Case-Oriented Selection:} Table~\ref{tab:tech_usecase_guidelines} summarizes indicative mappings between representative ISAC use cases and preferred antenna/waveform combinations, highlighting the underlying rationale. This complements the descriptive discussions in Tables~\ref{tab:isac_waveforms} and~\ref{tab:advanced_antennas} by emphasizing \emph{which} technologies are most appropriate under specific operational constraints and \emph{why}.
\end{itemize}

\begin{table*}[!ht]
	\centering
	\caption{Indicative Mapping of ISAC Technologies to Representative Use Cases}
	\label{tab:tech_usecase_guidelines}
	\renewcommand{\arraystretch}{1.35}
	\setlength{\tabcolsep}{8pt}
	\begin{tabular}{|p{2.5cm}|p{3cm}|p{4cm}|p{6cm}|}
		\hline
		\rowcolor[HTML]{E6E6E6}
		\textbf{Use Case} &
		\textbf{Preferred Antenna Configuration} &
		\textbf{Preferred Waveform(s)} &
		\textbf{Rationale} \\
		\hline
		\hline
		Autonomous driving / V2X &
		Massive MIMO at roadside/base stations, optional RIS panels at intersections &
		CP-OFDM for cellular links, FMCW or OFDM--FMCW hybrids for high-resolution radar, OTFS in extreme Doppler scenarios &
		Requires centimeter-level localization and high-throughput connectivity under high mobility. Massive MIMO provides angular resolution and multiplexing, FMCW/hybrid waveforms deliver fine range/velocity sensing, while CP-OFDM/OTFS maintain robust communication in fast-varying channels. \\
		\hline
		Industrial automation / smart factory &
		Dense small-cell massive MIMO, wall/ceiling-mounted RIS for NLOS coverage &
		CP-OFDM or SC-CP for robust links, hybrid FMCW–OFDM for precise ranging in robotic cells &
		Environment is cluttered but mostly low-to-moderate mobility. Massive MIMO plus RIS improves coverage and blockage robustness, while hybrid waveforms enable millimeter-level ranging for robots, SC-CP reduces PAPR for uplink from power-limited devices. \\
		\hline
		Indoor localization / human sensing (e.g., gesture, occupancy) &
		Moderate-size MIMO at APs, RIS to illuminate shadowed areas &
		CP-OFDM with sensing overlays, spread-spectrum or frequency-hopping ISAC where robustness to interference or privacy is critical &
		Mobility is moderate, but multipath-rich indoor channels dominate. Standard OFDM grids simplify reuse of Wi-Fi/6G infrastructure for sensing, while RIS and spread/frequency-hopping waveforms help mitigate interference and enhance coverage for fine-grained human sensing. \\
		\hline
		Wide-area mapping / environmental monitoring &
		Macro-cell massive MIMO, optional multistatic nodes with RIS-assisted paths &
		CP-OFDM with dual-domain ISAC overlays, spread-spectrum or adaptive waveforms for low-SNR, long-range sensing &
		Emphasis is on coverage and robustness over large areas rather than ultra-high data rates. Massive MIMO and multistatic architectures enable spatial diversity, while OFDM-based and spread-spectrum/adaptive waveforms provide flexibility to balance sensing range, resolution, and coexistence. \\
		\hline
		Battery-powered IoT ISAC nodes &
		Small MIMO or single-antenna devices, leverage network-side massive MIMO / RIS &
		SC-CP or narrowband spread-spectrum ISAC waveforms &
		Strict energy and hardware constraints at the device side favor low-PAPR, low-complexity waveforms. Most sensing processing is offloaded to infrastructure, which can employ massive MIMO/RIS for improved sensitivity while keeping terminal design simple. \\
		\hline
	\end{tabular}
\end{table*}

\subsection{Sensing-Aware Resource Allocation}
Sensing-aware resource allocation is an essential element in ISAC systems, particularly in 6G networks, where the problem lies in balancing frequently conflicting objectives regarding communication throughput and sensing precision. In order to improve system performance, efficient allocation requires management of multi-dimensional resources, such as time, frequency, space, code, polarization, and power, at the software and hardware layers \cite{3.3.1},\cite{3.3.2}. 

This resource allocation challenge can be posed under the same optimization formulation introduced in Section~\ref{FIIV}, Eq.~\eqref{eq:wsum}, i.e., maximizing the communication rate $R(\mathbf{W})$ under sensing-accuracy and power constraints. Specifically, the constraint on sensing precision is captured through a threshold on the CRB, and the precoding matrix $\mathbf{W}$ is subject to a total transmit-power budget. In this context, the problem formulation from Section~\ref{FIIV}, Eq.~\eqref{eq:wsum} directly applies, where $\epsilon$ represents the sensing precision threshold, and $P_{\max}$ signifies the highest allowed transmission power.

Recent studies have proposed integrated frameworks that encompass user fairness, stringent sensing QoS limits, and flexibility tailored to application needs. These frameworks utilize sophisticated mathematical techniques, including Pareto optimization for effectively navigating trade-offs between communication and sensing efficiency, game theory for managing competitive resource-sharing situations, and multi-granularity resource pooling methods to improve scalability and flexibility in densely populated network deployments. This facilitates dynamic sensing-aware utilization of resources that addresses the requirements of sensing and communication in complex ISAC scenarios, including cooperative sensing involving several base stations and mobile users \cite{3.3.3},\cite{3.3.4}.

\subsection{Machine Learning and AI Technologies}
AI and machine learning (ML) are transforming ISAC by enabling adaptive systems that surpass traditional analytical models, particularly in non-linear scenarios where mathematical tractability is limited.

From an ISAC perspective, AI is not merely a generic \textquotedblleft complexity reduction\textquotedblright $\,$tool but directly addresses three structural limitations of conventional designs. First, joint sensing and communication operation creates a coupled, high-dimensional state space (channels, target states, hardware non-idealities, traffic patterns, and control actions) that quickly renders classical parametric models and handcrafted algorithms intractable, learning-based methods can instead approximate these complex input–output mappings directly from data. Second, ISAC frequently operates in regimes where accurate stochastic models for clutter, interference, and hardware distortions are unavailable or strongly environment-dependent, making purely model-based approaches brittle, whereas AI can learn robust inference rules from raw or lightly processed RF measurements. Third, many emerging ISAC tasks are inherently semantic (e.g., activity recognition, gesture and object classification, and intent detection from RF signatures) and thus fall outside traditional estimation-theoretic formulations that focus solely on delay, Doppler, and angle. Here, deep representation learning is uniquely suited to extract high-level features from complex sensing data. These characteristics explain why AI is particularly powerful for problems such as joint beam and waveform adaptation under unknown dynamics, RF-based human sensing and classification, and cross-layer control policies that must react to non-stationary environments.
	
However, their practical deployment in 6G imposes severe challenges. First, \emph{data collection costs} are prohibitive; acquiring high-fidelity, synchronized RF sensing data with precise ground truth annotations requires complex experimental setups involving calibrated measurement equipment, unlike readily available image or text datasets~\cite{4.2.2}. Second, \emph{model generalization} remains a bottleneck, as data-driven models trained on specific channel realizations often fail when deployment environments (e.g., indoor vs. outdoor, urban vs. rural) or hardware configurations change, a phenomenon known as distribution shift~\cite{3.4.1a}. Third, the \emph{real-time inference overhead} of deep neural networks frequently exceeds the sub-millisecond latency budgets required for 6G physical layer processing, especially for operations like beam tracking and channel prediction that must occur within each OFDM symbol duration~\cite{3.4.01}. To address these trade-offs, two distinct paradigms have emerged with specific applicable scenarios:
	\begin{itemize}
		\item \emph{Data-Driven AI}: This approach relies on massive datasets to learn end-to-end mappings without explicit physical modeling. It is most applicable to \textit{high-level semantic tasks}, such as human activity recognition, gesture classification, and object detection from RF signals, where analytical models are either intractable or unavailable. However, it suffers from poor interpretability and limited robustness to out-of-distribution data~\cite{3.4.02, 3.4.03}.
		\item \emph{Model-Driven AI}: This paradigm integrates domain knowledge (e.g., physical channel models, radar signal processing algorithms) into deep learning architectures. Techniques such as \emph{deep unfolding networks}, which embed iterative optimization algorithms as neural network layers, are particularly effective for \textit{physical layer estimation tasks}, including channel estimation, beam alignment, and target parameter extraction. This approach offers superior sample efficiency, interpretability, and robust generalization compared to purely data-driven methods~\cite{3.4.04, 3.4.1, 3.4.2}.
\end{itemize}

\subsubsection{Feasibility on Resource-Constrained Edge Devices}
Deploying AI models on resource-constrained edge devices (e.g., IoT sensors, user equipment, vehicular terminals) presents fundamental challenges due to limited energy budgets (often $<1~W$), memory constraints (tens of kilobytes of RAM), and computational capacity (tens to hundreds of millions of operations per second). Standard deep neural networks (DNNs) with millions of parameters require gigaflops of computation per inference, making them infeasible for battery-powered devices requiring continuous ISAC operation. To bridge this gap, \emph{Lightweight AI} research has developed compression and optimization techniques that dramatically reduce model complexity while preserving accuracy:
\begin{itemize}
	\item \textit{Model Quantization} reduces numerical precision from $32$-bit floating-point to $8$-bit or even binary representations, yielding $4-32\times$ reductions in memory and energy consumption. Recent work demonstrates that quantized neural networks can maintain $>95\%$ of full-precision accuracy for ISAC channel estimation tasks while enabling inference on embedded ARM Cortex-M processors~\cite{3.4.1.00, 3.4.1.0}.
	\item \textit{Network Pruning} removes redundant weights and neurons based on magnitude or importance metrics, achieving $50-90\%$ sparsity (i.e., $10-50\times$ fewer operations) with minimal accuracy loss. Structured pruning methods tailored for wireless signal processing maintain performance while meeting the stringent latency requirements of sub-6~GHz and mmWave ISAC systems~\cite{3.4.1.01, 3.4.1.02}.
	\item \textit{Knowledge Distillation} trains compact "student" models by transferring knowledge from large "teacher" networks, enabling $10-100\times$ size reduction. For ISAC applications, distilled models have been shown to achieve near-teacher performance for beam prediction and target classification while fitting within the memory constraints of IoT platforms~\cite{3.4.1.03}.
	\item \textit{TinyML Frameworks} (e.g., TensorFlow Lite Micro, Edge Impulse) enable deployment of lightweight models on microcontrollers with $<256~KB$ memory, facilitating distributed, privacy-preserving ISAC intelligence at the extreme network edge. Field demonstrations show that basic sensing tasks, such as presence detection, motion classification, and anomaly detection, can execute on MCU-based platforms with $<10~ms$ latency and $<50~mW$ power consumption, meeting the requirements for battery-operated IoT ISAC nodes~\cite{3.4.1.04, 3.4.1.05}.
\end{itemize}

Despite these advances, fundamental trade-offs remain: aggressive compression degrades performance on complex tasks (e.g., multi-target tracking), and ultra-low-power inference restricts model capacity. Hybrid architectures that offload complex inference to edge servers while retaining lightweight preprocessing on end devices represent a promising direction for scalable ISAC deployment~\cite{6.3.7}.

\subsubsection{Multi-Objective Optimization}
The fundamental challenge in ISAC lies in simultaneously optimizing communication throughput and sensing precision, objectives that often conflict. Conventional optimization techniques for these multi-objective problems are frequently computationally demanding and encounter difficulties due to the non-convexity characteristic of numerous situations in the real world.  Deep learning models, especially deep reinforcement learning and meta-learning, are demonstrating significant efficacy in addressing complex multi-objective resource allocation and waveform design challenges. Through learning of optimal policies using comprehensive real-time interactions or simulations, these AI models can dynamically modify power levels, subcarrier distributions, and beamforming vectors, substantially surpassing traditional optimization methods in speed, adaptability, and capacity to manage non-linear and highly dynamic environments \cite{3.4.1.1},\cite{3.4.1.2}.

\subsubsection{Signal Processing}
AI algorithms are transforming numerous signal processing tasks in ISAC, resulting in significant enhancements in detection, classification, and estimation efficacy. Deep neural networks can effectively execute identification of targets and estimation of parameters (e.g., range, velocity, angle) even in adverse conditions, including low signal-to-noise ratios, significant clutter, or when confronted with inaccurate channel modeling and hardware limitations. In addition to conventional sensing, machine learning algorithms can classify object categories, recognize human activities, and deduce environmental attributes from communication signals or radar echoes. This data-centric methodology enables ISAC systems to comprehend intricate relationships that are challenging to model analytically, resulting in enhanced resilience and precision across a greater range of operational contexts \cite{3.4.1},\cite{3.4.2.1}.

\subsubsection{Data Augmentation}
A major obstacle in implementing data-centric AI models for ISAC is the lack of adequately large and diverse datasets, particularly for intricate scenarios of real-world that encompass various interference patterns, environmental factors, and targets \cite{3.4.3.1}. Generative AI methodologies, including generative adversarial networks (GANs) and diffusion models, provide an efficient solution to this problem by enabling substantial data augmentation. These models can generate high-fidelity synthetic data, including supplementary CSI samples, sensor readings, or radar echoes, that closely resemble data from the real world. The ability to produce diversified and genuine training data addresses the constraints of physical data collection, facilitating the development of more resilient and adaptable AI models for sensing, communication, and joint ISAC tasks, thus enhancing overall performance and reliability \cite{3.4.3.2}.

\subsubsection{End-to-End Learning}
The primary objective of numerous ISAC applications is to develop an autonomous system capable of optimizing its sensing and communication capabilities in a cohesive and unified manner \cite{3.4.4.1}. Reinforcement learning (RL) and multi-task learning frameworks are essential facilitators for this comprehensive optimization. Rather than optimizing sensing and communication independently, reinforcement learning agents can acquire ideal joint techniques through direct interaction with the environment, receiving rewards for attaining integrated communication throughput and sensing precision objectives. This enables ISAC systems to independently adjust to dynamic network conditions, environmental alterations (e.g., movable targets, varying clutter), and unforeseen occurrences. End-to-end learning aims to develop optimal policies that connect low-level signal processing with high-level network decisions, thereby enabling the creation of smart and self-optimizing ISAC systems that maximize overall system utilization across both domains \cite{3.4.4.2}.

At the same time, deploying AI in ISAC introduces critical challenges that go beyond generic “black-box” concerns and must be treated with particular rigor in 6G. Safety-critical applications such as autonomous driving, industrial automation, and e-health require formally verifiable performance guarantees, robustness against adversarial perturbations and spoofing, and graceful degradation under distribution shifts, whereas most existing learning-based ISAC models are validated only empirically on limited datasets. Embedding AI deeply into physical-layer sensing and beam control also raises new data governance and privacy issues, since RF sensing measurements can reveal sensitive information about users and environments even when no payload data is decoded. Furthermore, lifecycle management of AI models, including continuous retraining, on-line adaptation, certification across hardware and software updates, and standard-compliant testing, becomes a first-class engineering challenge tightly coupled to ISAC standardization and prototyping. Fully exploiting the benefits of AI in ISAC therefore requires not only algorithmic innovation, but also rigorous validation, monitoring, and standardization frameworks that are specifically tailored to learning-enabled, safety-critical RF systems.

\section{Current Trends} \label{CT}
Industry and standardization communities are increasingly concentrating on ISAC, which is emerging as a fundamental capability for 6G and beyond. Notable initiatives leading research and current trends of ISAC include:

\subsection{Localization, Mapping, and Environment Awareness}
Bringing together localization, mapping, and environmental sensing into a unified spatial intelligence platform is essential for creating smart and self-operating 6G services, marking an important trend with major impacts and difficulties. Simultaneous localization and mapping (SLAM) using ISAC combines RF sensing methods to find out where a device is and to create a map of the area around it, allowing it to work well in complex and changing environments where separate systems can't. Future initiatives focus on combining different types of sensors, like RF sensing with light detection and ranging (LIDAR), visual, and inertial data, to make systems more reliable and better at understanding their surroundings. Nonetheless, obstacles remain in harmonizing diverse sensing modalities, managing high-dimensional data streams in real time, and ensuring accuracy amidst mobility and environmental fluctuations. This basic understanding of space is important for key uses like self-driving cars, robot communication, and virtual reality, which require improvements in how algorithms can grow, how we represent meaning, and how we model the environment. Fixing these ongoing problems is important for moving ISAC from separate sensing and communication tasks to a complete network-wide smart system that meets 6G objectives \cite{2.2.4},\cite{4.1.1},\cite{4.1.2}.

\begin{figure*}[ht]
	\includegraphics[width=\linewidth,height=0.5\textheight]{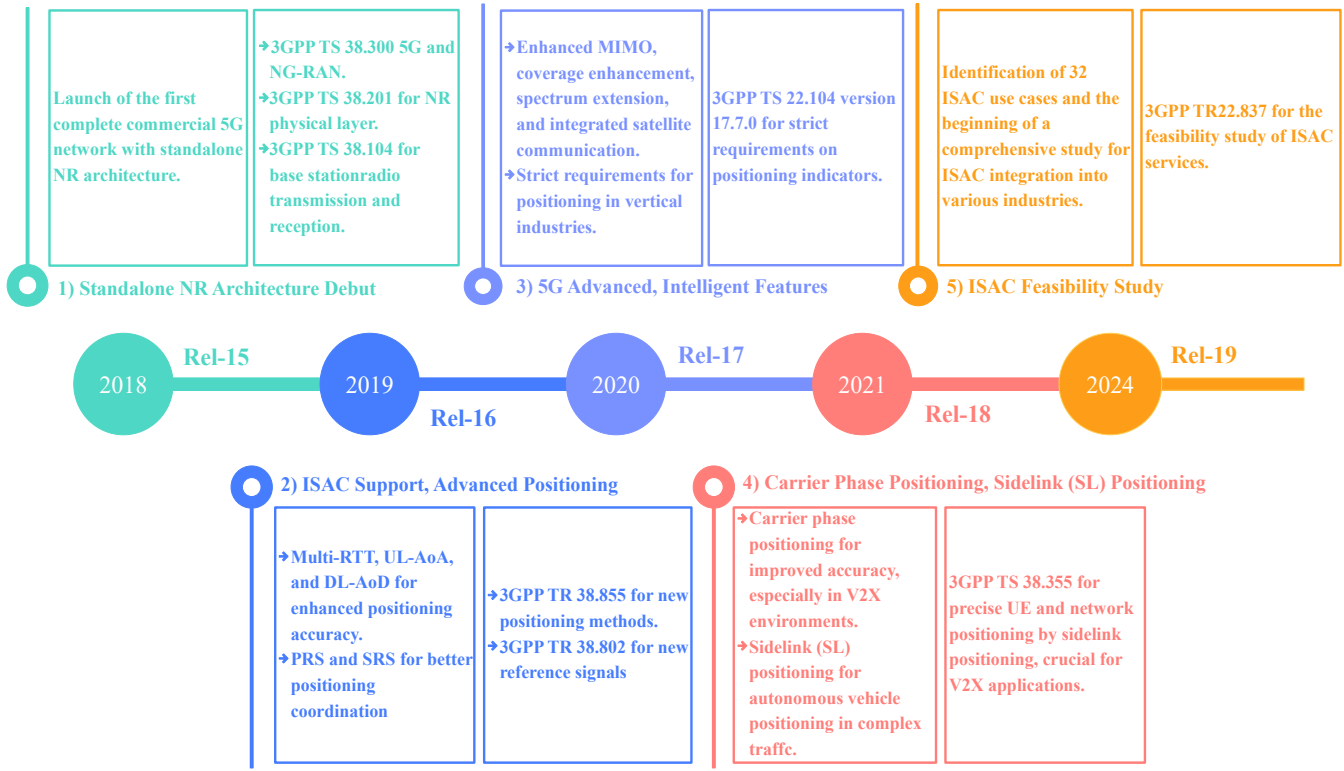}
	\centering
	\caption{Summary of 3GPP Advancements toward ISAC Standardization.}
	\label{fig:3GPP1}
\end{figure*}
\subsection{Industry Initiatives, Standardization, and System Prototyping}
The progress of ISAC is driven by synchronized global standardization efforts linked with practical prototype activities, which collectively address essential concerns, including interoperability, performance assessment, hardware viability, and security. Standards organizations such as 3GPP (commencing with Release 18 and the subsequent releases developed in Release 19, see Fig.~\ref{fig:3GPP1})\cite{3GPP}, along with the European Telecommunications Standards Institute (ETSI) Industry Specification Group (ISG) ISAC, have established fundamental architectures and key performance indicators (KPIs) that are important for ISAC functions. However, significant deficiencies persist in the definition of multifunction cross-layer protocol stacks, multitier integration frameworks, and universal channel and sensor models. 

Industry-driven momentum, led by entities such as the IEEE 802.11bf Task Group, ITU, and the 6G Infrastructure Association (6G-IA), has accelerated the adoption of ISAC by integrating sensing functionalities into Wi-Fi and cellular norms, facilitating applications ranging from environmental monitoring to smart home occupancy detection \cite{4.2.1}. Telecommunications firms such as Huawei, Nokia, Ericsson, and ZTE are rapidly advancing ISAC's vision through their 6G initiatives, transforming base stations into multifunctional sensing-communication hubs. 

In prototyping, adaptable platforms such as integrated RF front-ends, RIS, and software-defined radios (SDRs) are crucial for evaluating concepts and investigating design alternatives regarding hardware constraints, waveform intricacy, and energy consumption, particularly in domains like analog/digital conversion, metasurface antenna design, and RF chain integration, where size, cost, and scalability require meticulous equilibrium. 

Moreover, integrating suitable privacy and security protocols within the sensing-communication layers is an ongoing field of investigation \cite{onur1},\cite{T38},\cite{4.2.1a}, particularly in hazardous environments. Collective efforts in experimental prototyping, ecosystem coordination, and standards development emphasize the essential collaboration between technology suppliers, industry groups, and academic institutions to transform ISAC from an encouraging research domain into a reliable, scalable, and widely adopted technology vital for commercial 6G networks \cite{4.2.2},\cite{4.2.3},\cite{4.2.4}.

To further assist the reader in distinguishing the roles of these initiatives, Table~\ref{tab:standardization_comparison} provides a comparative analysis of their respective architectures, target deployment scenarios, and key performance indicators (KPIs).
	
	\begin{table*}[!ht]
		\caption{Comparative Analysis of Major Industry Standardization Initiatives for ISAC}
		\label{tab:standardization_comparison}
		\centering
		\renewcommand{\arraystretch}{1.35}
		\setlength{\tabcolsep}{8pt}
		\begin{tabular}{|p{3cm}|p{4.0cm}|p{4.5cm}|p{4.0cm}|}
			\hline
			\rowcolor[HTML]{E6E6E6}
			\textbf{Standard / Body} & \textbf{Primary Focus \& Deployment Role} & \textbf{System Architecture} & \textbf{Target KPIs \& Capabilities} \\
			\hline
			\hline
			\textbf{IEEE 802.11bf} (Wi-Fi Sensing) \cite{4.2.5} &
			\textbf{WLAN / Indoor:} Focuses on license-exempt bands (2.4/5/6/60 GHz) for short-range, consumer-grade sensing such as home monitoring and gesture control. &
			\textbf{Interface-Centric:} Defines MAC/PHY protocols for channel measurement setup. Supports Monostatic, Bistatic, and Multistatic topologies without centralized core control. &
			Targets human presence detection with approx. 100 ms latency and room-scale range ($<20$ m). Key metrics include motion detection, breathing rate, and gesture recognition. \\
			\hline
			\textbf{3GPP Rel-19/20} (Cellular ISAC) \cite{4.2.6} &
			\textbf{Cellular / Wide-Area:} Focuses on licensed bands (FR1/FR2) for outdoor, high-mobility (V2X), and critical infrastructure sensing. &
			\textbf{Network-Centric:} gNB-integrated sensing coordinated by the 5G Core (5GC). Architectures include Monostatic (gNB-only) and Cooperative (Multistatic) sensing. &
			Targets safety-critical sub-10 ms latency with high precision ($<1$ m outdoor, $<10$ cm indoor). Tightly integrated with URLLC and eMBB services for V2X applications. \\
			\hline
			\textbf{ETSI ISG ISAC} \cite{1.21} &
			\textbf{Framework Definition:} Pre-standardization body defining generic use cases, channel models, and evaluation methodologies. &
			\textbf{Service-Oriented:} Defines a reference architecture focusing on cross-layer interfaces and \textquotedblleft Sensing as a Service\textquotedblright $\,$capability exposure. &
			Focuses on defining standardized evaluation metrics such as detection probability ($P_d$), false alarm rate ($P_{fa}$), and sensing update rate, rather than specific performance values. \\
			\hline
		\end{tabular}
	\end{table*}

\subsection{Expanded Use Cases and Applications}
Future research directions are shaped by the various emerging applications facilitated by ISAC, which are crucial to the 6G vision of pervasive intelligence. Each of these applications poses unique operational and technical challenges. Intelligent transportation systems (ITS) necessitate highly reliable, low-latency, sub-centimeter localization and collaborative perception in rapidly changing, hazardous environments \cite{4.3.0}, which require ISAC architectures to dynamically manage sensing-communication trade-offs while adhering to strict timing guarantees \cite{3.4.1}. Human-centric sensing applications raise concerns regarding privacy and ambient intelligence, requiring innovative hardware-software co-designs that integrate privacy-preserving sensing into communication infrastructures. Intelligent manufacturing facilities and industrial IoT utilize ISAC for localization, automation, and robotics, yet face challenges from severe multipath, interference, and rigorous reliability requirements. Immersive technologies such as XR and tactile internet show how different types of communication can work together, where smart ISAC communication reduces unnecessary data while keeping the environment accurate \cite{XR}. Every domain emphasizes the necessity for adaptable, context-sensitive ISAC solutions that can address diverse QoS and sensing precision demands, thus prompting investigation into application-specific waveform tailoring, resource distribution, and AI-driven policy adaptation \cite{2.2.2},\cite{4.3.1},\cite{4.3.2}.

\subsection{New Spectrum and Hardware Paradigm}
The use of innovative spectral domains, such as THz and mmWave bands, represents a significant advancement for ISAC by delivering enormous bandwidths that enable micron-level sensing resolution alongside ultra-high throughput communication \cite{4.4.1}. Nonetheless, utilizing these frequency bands has inherent challenges: significant propagation losses, limitations in hardware design, and blockage sensitivity place rigorous demands on transceiver structures and signal processing techniques. Resolving such challenges requires a cooperative integration of sophisticated beamforming, full-duplex operation, and RIS-enabled programmable radio environments, which together enable dynamic control over the environment and enhance spectrum efficiency \cite{4.4.2}. Cell-free massive MIMO improves spatial diversity and extends coverage. However, its dispersed characteristics raise unresolved research challenges in synchronization, scalable coordination, and distributed resource management\cite{4.4.3}. In addition, evolving integrated RF chain designs and metasurface antennas have to meet the conflicting demands of power efficiency, compactness, and high-fidelity shared existence in sensing and communication. Confronting these issues requires interdisciplinary innovation, combining circuit design, electromagnetics, and predictive algorithms in order to develop hardware paradigms that align with the deployment and performance goals of 6G ISAC \cite{4.4.4}. Refer to Section~\ref{AT} for foundational antenna technologies.

\subsection{AI and Data-Driven ISAC Architectures}
Artificial intelligence is swiftly becoming the cornerstone for the development of intelligent, adaptable, and scalable ISAC systems in next-generation networks. Current advancements include real-time AI-driven control systems, including deep reinforcement learning and meta-learning, that independently enhance beamforming, resource allocation, and waveform adaptability in rapidly changing channel and environmental conditions \cite{3.4.1}. AI-driven semantic information extraction transforms raw radio frequency signals into comprehensible environmental insights, such as object trajectories and semantic maps, thereby enhancing decision-making and optimizing task-specific communication. Distributed AI frameworks, such as federated and edge learning, have arisen as viable solutions to privacy and latency issues by decentralizing intelligence across network nodes and user devices. However, significant research gaps remain about the cross-domain generalization, sample efficiency, and explainability of these learning models in diverse ISAC implementations. The integration of adaptive AI with sensing-aware management of resources is set to establish robust, self-optimizing ISAC networks that seamlessly combine communication and sensing functionalities for complex and essential 6G applications \cite{4.5.1},\cite{4.5.2},\cite{4.5.3}.

\subsection{Waveform and Signal Processing Advances}

The future of ISAC waveform design is evolving beyond traditional methods like OFDM, advancing toward innovative paradigms such as OTFS modulation, chirp-based waveforms, and complex hybrid formats. OTFS, in particular, offers significant benefits for delay-Doppler estimation and maintains robust performance in highly dynamic 6G environments, enabling superior integration of communication and sensing functionalities essential for next-generation wireless systems \cite{4.2.3},\cite{4.6.1}.

A primary challenge in this domain is balancing the trade-off between sensing precision, typically assessed by metrics such as the CRB, and transmission throughput, all while maintaining computational efficiency for real-time functionality. Achieving this balance is critical for ensuring that sensing tasks do not degrade the quality of service for communication users.

Building on these advances, next-evolution waveforms (NEW), such as orthogonal delay-doppler division multiplexing (ODDM) \cite{ODDM} and affine frequency division multiplexing (AFDM) \cite{AFDM}, are attracting significant attention for their ability to meet the stringent dual requirements of high-resolution sensing and high-throughput communication in future ISAC systems.

ODDM achieves its superior performance by mapping information symbols onto a finely sampled grid in the delay domain, effectively creating a two-dimensional orthogonal basis that spans both delay and Doppler dimensions. By employing a delay-doppler orthogonal pulse (DDOP) \cite{DDOP} shaping filter, ODDM attains near-ideal ambiguity function characteristics, namely, a sharp main lobe in delay and Doppler coupled with low sidelobe levels. This yield translates directly into improved target resolution and reduced mutual interference among reflected echoes, even in the presence of severe multipath and high relative velocities. On the communications side, the orthogonality across delay taps enables simple per-tap equalization and low-complexity receiver implementations. Furthermore, ODDM's inherent resilience to channel delay spread makes it particularly effective in rich scattering environments, such as urban vehicular or indoor millimeter-wave scenarios, where traditional multi-carrier schemes suffer from inter-symbol interference.

AFDM, by contrast, constructs its waveform by applying an affine (i.e., time-scaling and frequency-shifting) transform to a prototype chirp signal, thereby generating a family of linearly time- and frequency-diverse subcarriers. Each subcarrier exhibits a constant group delay characteristic and uniform Doppler tolerance, yielding a flat ambiguity surface that markedly simplifies joint delay-Doppler channel estimation. The chirp-based nature of AFDM ensures low PAPR relative to OFDM, which reduces transmitter amplifier back-off requirements and enhances power efficiency, an essential consideration for battery-constrained platforms such as unmanned aerial vehicles or wearable sensors. On the sensing front, AFDM's constant instantaneous frequency sweep provides a built-in matched filter structure that maximizes signal-to-noise ratio for range and velocity measurement. From a communications perspective, the affine transform framework admits straightforward integration of index modulation and error correction codes, enabling flexible trade-offs between spectral efficiency and link reliability.

Furthermore, the growing theory of semantic-aware communication focuses on the transmission of task-relevant semantic information instead of raw data, therefore considerably minimizing overhead and enhancing alignment with sensing objectives. Adaptive signal processing techniques, such as RIS-assisted beamforming and ambiguity function shaping, are essential for alleviating interference and spectrum conflicts. Techniques like directional modulation and S-IM-OFDM (refer to Section~\ref{OFDM}) demonstrate the increasing synergy between waveform innovation and unified sensing-communication architecture, emphasizing the synthesis of waveform, spatial control, and artificial intelligence to achieve resilient, high-performance ISAC in 6G and beyond \cite{4.6.2},\cite{4.6.3}.

\subsection{Benchmark Datasets and Testbeds}
The development of benchmark datasets and experimental testbeds for ISAC is a constantly advancing research domain essential for connecting theoretical progress with practical implementation in 6G networks. Despite considerable advancements in algorithmic and architectural design, the domain continues to encounter a deficiency of open-access, large-scale datasets that concurrently contain the complexities of co-designed sensing and communication signals alongside precise environmental ground reality. Until recently, the majority of research depended on synthetic datasets produced by simulators or proprietary hardware testbeds with restricted public access, limiting reproducibility and the ability to compare methodologies \cite{4.3.1},\cite{4.7.1}.

\subsubsection{Current Status and Key Initiatives}
Several pioneering efforts have begun to address this dataset gap by releasing multi-modal data collections that integrate communication parameters such as CSI with radar-like sensing echoes. Researchers have created synchronized CSI and echo datasets from mmWave ISAC prototype systems, utilizing realistic vehicular environments where targets and clutter coexist. These datasets enable systematic evaluation of the precision of joint localization alongside communication throughput under varying mobility and multipath conditions, facilitating the development of unified sensing-communication algorithms \cite{4.3.1}. Complementing raw signal datasets, annotation-enriched collections have emerged that support AI-driven ISAC methods. The compilation of AI-ready datasets containing semantic labels such as object categories and trajectory information is crucial for supervised and semi-supervised learning paradigms targeting intelligent ISAC solutions. Such data foundations are instrumental in the advancement of semantic sensing, environment-aware communication, and cross-modal data fusion in practical 6G contexts. Community-driven repositories and effort groups are also beginning to catalog these resources to encourage broader uptake and benchmarking standards \cite{4.7.1.1}.

\subsubsection{Experimental Testbeds}
On the hardware front, experimental testbeds have progressed from integrating discrete radar and communication systems to highly integrated platforms capable of real-time, simultaneous sensing and communication \cite{4.7.2.1a},\cite{4.7.2.1b}. The introduction of wideband mmWave ISAC prototypes utilizing large antenna arrays with fully digital beamforming to dynamically track multiple targets while maintaining communication links was validated in live urban field trials. These testbeds quantify crucial trade-offs between sensing resolution and data rate under realistic interference and mobility, providing important lessons for system design \cite{4.7.2.1}. Advances in programmable wireless environments through RIS have been explored extensively. The RIS-enhanced ISAC platforms dynamically modify propagation channels via controllable metasurface elements, resulting in improved sensing fidelity and communication reliability across different spatial locations. Their experimental evaluations demonstrate how RIS can mitigate multipath fading and extend sensing coverage in complex scenarios \cite{4.2.1, ISACMETA}. 

For application-driven contexts such as autonomous driving and indoor positioning, multi-sensor ISAC platforms have been developed. Vehicular and indoor localization testbeds are developed that integrate with RF sensing, inertial measurement units, and other modalities to create synchronized, ground-truthed datasets encompassing both communication performance and high-accuracy spatial measurements. These testbeds serve as benchmarks for multi-modal ISAC fusion algorithms essential for robust autonomous systems and immersive environments \cite{4.7.2.2},\cite{4.7.2.3}.

\subsubsection{Challenges and Outlook}
In spite of considerable advancements, crucial challenges persist in ISAC benchmark datasets and testbeds, notably the lack of standardized protocols that concurrently assess sensing and communication performance metrics, including detection probability, localization error, latency, throughput, CRB, and bit error rate (BER), hindering consistent comparison across various research studies \cite{4.7.3.1}. Furthermore, there is a significant deficiency of extensive, open-access datasets that include diversified environments, multi-user scenarios, and mobility patterns, which constrains the robustness and generalization of algorithms. The generation of such datasets is intricate because of the requirement for coordinated RF measurements, environmental annotations, and ground reality in unpredictable circumstances. Practical testbeds encounter engineering challenges concerning real-time processing, calibration, scalability, and hardware complexity, necessitating innovative cross-layer design and optimization to integrate sensing and communication without compromising the introduction of interference or compromising power efficiency \cite{4.7.3.2},\cite{4.7.3.3}. Furthermore, joint efforts among standards organizations, industry, and academia to develop standardized KPI frameworks, facilitate open data sharing, and design adaptable, modular testbed platforms are crucial \cite{4.2.1},\cite{4.3.1}. Advances in the use of AI for transfer learning, data augmentation, and annotation present promising opportunities to enhance dataset utility and model generalization, thus expediting ISAC's shift from theoretical frameworks to scalable 6G applications such as immersive XR, autonomous vehicles, and smart manufacturing experiences \cite{3.4.1},\cite{3.4.2},\cite{4.7.3.4}.

\section{Research Challenges and Open Problems} \label{RCOP}
Leveraging the potential of ISAC in 6G networks necessitates addressing various significant research challenges and open problems that hinder the practical, efficient, and secure deployment of the system.

\subsection{Waveform Deployment Challenges}
Despite their promise, practical deployment of ODDM and AFDM entails several open challenges. A precise design for shaping pulses and chirp parameters must simultaneously balance sidelobe suppression against filter length, computational burden, and hardware resource constraints. Pilot and synchronization schemes require innovative two-dimensional pilot patterns to adapt to the unique delay-Doppler and affine waveform structures and to avoid pilot contamination in dense multi-user ISAC networks. Finally, real-time hardware implementations on digital signal processors and field-programmable gate arrays demand careful algorithm-architecture co-design and low-overhead parallelization to achieve the low latency and high throughput necessary for seamless communication–sensing coexistence. Addressing these challenges will be critical to fully realizing the potential of NEW waveforms in next-generation ISAC systems \cite{4.3.1}\cite{ODDM}\cite{AFDM}.

\subsection{Coexistence and Interference Management}
A key challenge in implementing ISAC systems for 6G networks is guaranteeing the efficient coexistence of concurrent sensing-communication operations within identical spatial, temporal, or spectral resources. This dual-purpose operation results in unnecessary mutual interference, especially as both functions may overlap or multiplex on the same frequencies and hardware. Managing this kind of interference beyond fundamental spectrum distribution necessitates innovative scheduling and adaptive signal processing techniques capable of dynamically sensing, predicting, and mitigating interaction between the two domains. Innovative solutions, such as joint waveform optimization, RIS-assisted spatial filtering, and adaptive beamforming, can considerably minimize interference while simultaneously increasing system complexity and design trade-offs \cite{5.1.1},\cite{5.1.2}. The development of reliable coexistence protocols, particularly in densely populated environments with numerous ISAC transceivers operating in close vicinity, continues to be an active research domain, emphasizing environment-adaptive waveform design, interference-aware scheduling, and machine learning-based coordination of ISAC nodes\cite{5.1.3}.

\subsection{Hardware Constraints}
The main development consists of employing shared hardware components to concurrently execute sensing and communication functions, thereby attaining efficient dual functionality without compromising the performance of either capability. Nevertheless, practical limitations, such as restricted dynamic range, phase noise, suboptimal calibration, and amplified non-linearity, can introduce coupling that diminishes overall system efficacy. Wideband operation at mmWave and THz frequencies significantly challenges the capabilities of advanced mixed-signal and analog circuitry, necessitating highly integrated, power-efficient, and reconfigurable front-end designs. The joint design of adaptable transceiver architectures needs to balance high communication throughput with the accuracy necessary for comprehensive sensing while adhering to stringent energy, cost, and size limitations. This hardware constraint is particularly evident in emerging contexts such as cell-free and full-duplex deployments, where concurrent transmission and reception, along with distributed architectures, intensify non-idealities and synchronization challenges. Ongoing advancements in compact metasurface components, calibration algorithms, and RF design are essential to unlock the full potential of ISAC technology for forthcoming 6G devices \cite{3.4.1},\cite{5.2.1},\cite{5.2.2}.

\subsection{Standardization Gaps}
Despite considerable progress achieved across numerous international and industry-driven initiatives, the standardization framework for ISAC remains disconnected and insufficient. Fundamental challenges remain regarding developing integrated protocol designs and performance metrics that effectively combine the conventionally separate sensing and communication domains. The variation in channel modeling methodologies, sensing techniques, and system architectures prevented standard interoperability and vendor-agnostic implementation across various network environments. The lack of robust privacy and security standards specifically designed for pervasive sensing data impedes adoption, especially in sensitive human-centered applications. Addressing these gaps requires dedicated research on aligning diverse specifications, creating universal benchmarking methods, and establishing adaptable yet resilient cross-layer interfaces. This necessitates improved collaboration among standard organizations, industry consortia, and academic institutions to develop adaptable, future-compatible standards that can progress with constantly evolving ISAC technologies and applications \cite{3.4.1},\cite{4.3.1},\cite{5.3.1}.

\subsection{Tradeoffs Between Accuracy, Reliability, and Spectral Efficiency}
The challenge of balancing communication reliability, sensing accuracy, and spectral efficiency persists as significant in dynamic 6G networks, where applications have different QoS needs \cite{4.7.3.2},\cite{5.4.1}. For instance, giving more spectrum to improve sensing resolution could imply that there is less bandwidth for communication and vice versa.  Similarly, if ultra-reliable communication is of greater significance than high-fidelity sensing in mission-critical situations, the resources needed for high-fidelity sensing may be limited. To attain a Pareto-optimal balance between these goals, one requires advanced multi-objective optimization techniques that often use machine learning, AI,  or game-theoretic mechanisms to adaptively distribute resources based on changing network needs \cite{5.4.2}. Uncertainties in the environment, mobility, and channels that change over time make these factors even more challenging to model accurately. Prospective research should focus on developing smart, flexible systems that can learn and adjust to changing conditions, looking into semantic-aware communication to cut down on unnecessary data transmission, and verifying proposed solutions in realistic, various 6G testbeds \cite{5.4.3}.

\subsection{Prioritization of Future Research Efforts}
While the challenges outlined above are multifaceted, they carry varying degrees of urgency for the successful rollout of 6G. To assist researchers in prioritizing their efforts, Table~\ref{tab:challenge_prioritization} ranks these main challenges by their anticipated impact on deployment. We classify them into \textit{Critical Deployment Blockers} (immediate standardization and coexistence hurdles), \textit{Performance Scalability Enablers} (hardware and waveform optimizations), and \textit{Long-Term Ecosystem Trust} (security and privacy), thereby providing a structured roadmap for future investigation.
	
	\begin{table*}[!ht]
		\caption{Strategic Prioritization of ISAC Research Challenges}
		\label{tab:challenge_prioritization}
		\centering
		\begin{tabular}{|p{3cm}|p{3cm}|p{5.0cm}|p{5.0cm}|}
			\hline
			\rowcolor[HTML]{E6E6E6}
			\textbf{Priority Level} & \textbf{Key Challenge Area} & \textbf{Primary Obstacles \& Impact} & \textbf{Recommended Research Focus} \\
			\hline
			\hline
			\textbf{Priority 1: Critical Deployment Blockers} (Immediate) &
			\textbf{Standardization \& Interoperability} &
			Lack of unified channel models, cross-layer signaling protocols, and evaluation metrics prevents multi-vendor integration and mass commercial adoption. &
			Develop technology-agnostic interfaces in 3GPP Rel-20 \cite{3GPP}, Harmonize sensing KPIs (e.g., $P_d$, CRB) with comms KPIs (throughput) in standard specs. \\
			\cline{2-4}
			& \textbf{Coexistence \& Interference} &
			Mutual interference between sensing/comms functions and legacy systems causes severe reliability degradation, rendering the system operationally unviable. &
			Design robust self-interference cancellation (SIC) circuits, Develop dynamic spectrum sharing (DSS) algorithms and localized null-steering precoders \cite{5.5.1}. \\
			\hline
			\textbf{Priority 2: Performance Scalability} (Short-to-Mid Term) &
			\textbf{Hardware Constraints} &
			Non-linearities, power amplifier (PA) efficiency, and limited isolation in compact transceivers restrict sensing range and resolution, limiting the realizable gain of ISAC. &
			Investigate hardware-impairment-aware waveform design, Develop low-cost, high-isolation full-duplex antenna architectures. \\
			\cline{2-4}
			& \textbf{Waveform Optimization} &
			Conventional OFDM suffers from high PAPR and autocorrelation sidelobes, limiting sensing accuracy in power-constrained wideband scenarios. &
			Optimize joint waveforms (e.g., OTFS, AFDM) for ambiguity function shaping \cite{4.6.2}, Explore sensing-centric pilot placement strategies. \\
			\hline
			\textbf{Priority 3: Optimization \& Trust} (Long Term) &
			\textbf{Privacy \& Security} &
			Ubiquitous sensing raises severe privacy concerns (surveillance risks) and expands the attack surface (e.g., sensing data spoofing). &
			Formulate \textquotedblleft Privacy-by-Design\textquotedblright $\,$sensing protocols \cite{5.5.3}, Research physical layer security (PLS) and jammed-resilient sensing algorithms. \\
			\hline
		\end{tabular}
	\end{table*}

\section{Design Recommendations and Insights} \label{DRI}
The effective design for next-generation wireless systems is based on a comprehensive approach that incorporates flexible architectures, interdisciplinary optimization, and alignment with varied application requirements \cite{3.4.2.1},\cite{4.2.2}. Environmental awareness must be integrated into communication functions from the initial design phases, facilitating context-sensitive operation in dynamic deployment contexts \cite{3.4.2}. Flexibility should encompass both software and hardware domains, where modular architectures and scalable interfaces provide adaptation to shifting standards and applications. Instead of addressing components of the system independently, efficient coordination, encompassing waveform strategies, signal processing, and resource allocation, should focus on robustness, energy efficiency, and long-lasting scalability in the context of upcoming 6G technologies \cite{6.1, 6.2, 6.3}. The subsequent subsections contain broad perspectives on architectural structuring, system-level integrating approaches, and application-driven guidance.

\subsection{Architectural Guidelines}
To provide effective implementation in 6G networks, architectural mechanisms must facilitate diversified communication methods and environmental awareness using loosely linked but compatible modules \cite{4.3.2}. A layered design architecture that segregates essential functions like signal processing, resource management, and waveform generation improves implementation adaptability while facilitating distinct interfaces for inter-layer collaboration \cite{5.3.1}. Hierarchical architectures are particularly endorsed, wherein edge computing layers facilitate low-latency integration of sensing and communication data, releasing computational constraints on user equipment and positioning intelligence nearer to the surroundings \cite{6.1.1},\cite{6.1.2}. Distributed antenna systems and RIS components can operate as hybrid nodes, transmitting information that is processed through software-defined interfaces. The implementation of standardized APIs and shared hardware platforms, like common transceiver modules and reconfigurable analog front-ends, enhances hardware reutilization and cost-effectiveness \cite{5.1.1},\cite{6.1.3}. Furthermore, security and privacy should be fundamental, incorporating specialized isolation procedures for critical sensing data alongside conventional communication protection methods \cite{3.4.1},\cite{6.1.4}.

\subsection{Best Practices for Joint System Design}
High-performance integrated systems gain advantages from optimization strategies that consider resource management, beamforming, and waveform design as interconnected factors within a multi-objective context \cite{3.4.1.1},\cite{6.2.1}. Balancing conflicting objectives, such as communication reliability, sensing accuracy, and spectral efficiency, necessitates adaptable and context-sensitive techniques that respond to real-time dynamics in the environment and channel \cite{5.2.2},\cite{6.2.2}. Adaptive beamforming, robust channel coding, and dynamic waveform selection are essential facilitators \cite{6.2.3}. The integration of AI-powered control engines, trained on multi-modal data, facilitates low complexity in making decisions and consistent performance across diverse scenarios \cite{6.2.4}. A modular signal processing architecture must provide simultaneous sensing and communication operations to streamline upgrades and enhance prototyping \cite{3.2.2.2},\cite{4.5.2}. Efficiency in energy usage continues to be a design imperative at both algorithmic and hardware stages, especially for edge-based and distributed systems, to guarantee sustainable development as the system evolves \cite{3.4.2},\cite{6.2.5}.

\subsection{Use-Case-Specific Guidance}
Context-aware implementations are crucial to address the specific needs of various 6G applications \cite{6.3.1},\cite{6.3.2}. Ultra-low latency and high reliability are required for intelligent transportation and automotive systems, with a focus on precise localization as well as robust connectivity in contexts where safety is crucial \cite{6.3.3},\cite{6.3.4}. Industrial automation necessitates architectures that are equipped with distributed intelligence to facilitate fault-tolerant control, capable of accurate synchronization, and resistant to interference \cite{6.1.2},\cite{6.3.5},\cite{6.3.6}. Human-centric applications, such as remote health monitoring, smart environments, and gesture recognition, require low-profile hardware, adaptive systems, and robust privacy protections that balance data utility with communication load \cite{2.2.2},\cite{6.3.7},\cite{6.3.8}. Immersive sectors like the tactile internet and extended reality necessitate semantically aware resource planning and effective sensor fusion for context-aware, smooth interaction \cite{XR},\cite{6.3.9}. In all sectors, iterative development cycles and active stakeholder participation are essential for attaining operational readiness and real-world feasibility \cite{4.2.1},\cite{6.3.10},\cite{6.3.11},\cite{6.3.12}.

\section{Future Perspectives} \label{FP}
Realizing a new epoch of wireless connectivity necessitates going beyond conventional communication paradigms to adopt systems that adeptly analyze and leverage environmental information. This innovation allows networks to transmit data while also extracting significant context, thereby improving efficiency, adaptability, and the range of applications \cite{3.4.2.1},\cite{4.7.2.2}. However, as ISAC systems become increasingly integrated with critical infrastructures, ensuring robust security against emerging threats that simultaneously target both sensing and communication functionalities will be paramount for their successful deployment \cite{3.4.1, 4.3.2}. The combination of advanced semantic understanding, quantum technology, and integration with future network paradigms is poised to shape wireless services and their impact on society \cite{7.1},\cite{7.2}.

\subsection{Beyond 6G: Semantic ISAC and Quantum-Enabled ISAC} 

\subsubsection{Semantic ISAC}
Beyond the 6G horizon, developing concepts focus on networks that evolve from conventional data transmission to intelligent systems capable of retrieving and responding to information pertinent to the task \cite{7.1.1}. The upcoming paradigm emphasizes semantic communication, wherein the network understands operational context and user intent within sensing and communication processes. This facilitates applications including immersive extended reality, smart healthcare, and cognitive robotics by emphasizing expert knowledge over raw data streams.

\textbf{Technological Readiness \& Development Roadmap:} 
Within the current IMT-2030 (6G) vision, Integrated Sensing and Communication and Integrated AI and Communication are recognized as core usage scenarios, while \emph{Semantic ISAC} should be regarded as an early-stage research direction that builds on these pillars rather than as a fully specified feature \cite{7.1.1a}. From today’s perspective, Semantic ISAC remains at a low technology readiness level (roughly TRL 2–3), with most results limited to conceptual models, simulations, and small-scale prototypes. In order to align with the IMT-2030 timeline, which targets completion of 6G specifications around 2030, the evolution of Semantic ISAC can be organized into a phased, hierarchical roadmap as follows:
\begin{itemize}
	\item \textit{Phase 1 – Constrained Semantic ISAC in Vertical Pilots (Pre-IMT-2030 Standard Finalization):} 
	In the near term, semantic ideas are expected to be evaluated in constrained and well-structured scenarios, such as industrial robotic control, smart factories, and tightly controlled sensing environments. In these use cases, the semantic vocabulary (e.g., “object present/absent,” “risk level high/medium/low,” “target trajectory deviation”) can be predefined, and the impact of semantic compression on sensing accuracy and communication reliability can be systematically assessed. This phase primarily targets research testbeds and pre-standardization activities, where semantic encoders/decoders are layered on top of existing ISAC waveforms and evaluated against classical bit-wise baselines.
	\item \textit{Phase 2 – Semantic-Native ISAC Functions in 6G Evolution (Around and Beyond Initial IMT-2030 Deployments):} 
	In a subsequent phase, and broadly aligned with later 6G releases (e.g., 3GPP Release 20 and beyond), semantic concepts are expected to be increasingly co-designed with ISAC at the physical and MAC layers. Rather than operating as an overlay, semantic ISAC will require:
	\begin{itemize}
		\item \emph{Task-Oriented Semantic Metrics:} A key theoretical challenge is to move from bit-wise metrics to task-level metrics that capture how much of the transmitted information is truly relevant for the sensing and control objective. One research direction is to define a notion of “semantic entropy” that distinguishes task-relevant information from task-irrelevant redundancy, providing a bridge between classical Shannon entropy and task-oriented performance. A widely accepted and rigorously validated definition of such metrics is, however, still missing.
		\item \emph{Semantic-Native Waveform and Resource Design:} Existing ISAC waveforms (e.g., CP-OFDM, OTFS, chirp-based schemes) are optimized primarily for reconstructing physical parameters such as delay, Doppler, and angle. Semantic ISAC, by contrast, aims to allocate time, frequency, and space resources according to information importance at the task level. This calls for joint design of modulation, coding, and beamforming strategies that directly weight more critical semantic features (e.g., “imminent collision” vs. “background clutter”), rather than treating all bits equally.
		\item \emph{Semantic-Aware Joint Sensing–Communication Optimization:} Classical ISAC trade-offs are expressed in terms of rate versus sensing error bounds (e.g., CRB). Semantic ISAC will require new formulations where the objective is task performance (e.g., probability of correct decision, detection latency) subject to constraints on sensing and communication resources. This demands new optimization frameworks and performance bounds that explicitly account for semantic relevance.
	\end{itemize}
	\item \textit{Phase 3 – Network-Wide Semantic ISAC (Post-IMT-2030 Horizon):}
	In the longer term, Semantic ISAC is envisioned to evolve towards network-wide semantic coordination, where multiple ISAC nodes and base stations exchange and fuse semantic information rather than raw measurements. Edge intelligence and federated learning will play a crucial role in enabling distributed maintenance of semantic models that adapt to local environments (e.g., different traffic patterns, indoor vs. outdoor sensing conditions) while maintaining global consistency. At this stage, semantic processing becomes an integral part of the end-to-end design: from waveform generation and sensing inference up to application-level decision making.
\end{itemize}

Overall, while the IMT-2030 and 6G-IA roadmaps do not yet specify a dedicated timeline for Semantic ISAC as a separate feature, the above hierarchical view is consistent with their broader vision: initial 6G specifications providing the foundational ISAC and AI capabilities, and subsequent releases gradually introducing more task- and semantics-oriented functionality.

\subsubsection{Quantum-Enabled ISAC}

The simultaneous integration of quantum technologies presents transformational potential \cite{7.1.2, 7.1.3}. Quantum sensing methodologies and secure protocols for communication centered on entanglement may result in fundamentally novel interaction paradigms, high-assurance security, and exceptionally precise environmental assessments in the real world \cite{7.1.4, 7.1.5}. Proof-of-concept implementations utilizing photonic systems and quantum-enhanced radar-like schemes indicate that these advancements may ultimately transform network perception and response \cite{7.1.6, 7.1.7}.

\textbf{Technological Readiness \& Development Roadmap:} 
In contrast to Semantic ISAC, Quantum-Enabled ISAC is at a very early technical maturity, best characterized as a long-term research direction. Existing quantum sensing and communication demonstrations are mostly confined to laboratory environments with highly specialized hardware. In TRL terminology, this corresponds roughly to TRL 1–2 for mobile network grade quantum ISAC. Current 6G roadmaps from ITU-R and 6G-IA emphasize ISAC, AI, and security as key pillars, and also highlight quantum technologies as important enablers for trustworthiness and ultra-precise sensing in the longer term \cite{7.1.5a}. However, they do not foresee quantum ISAC as a mainstream capability within the initial IMT-2030 deployment timeframe, but rather as a post-6G evolution. A structured, hierarchical view of its development can be summarized as:
\begin{itemize}
	\item \textit{Phase 1 – Hybrid Quantum–Classical Sensing at Infrastructure Nodes (Long-Term 6G R\&D Stage):}
	In the first stage, quantum devices are expected to appear as auxiliary sensors at fixed infrastructure nodes (e.g., macro base stations, specialized sensing hubs) rather than being integrated into user equipment. Examples include Rydberg-atom–based receivers and other quantum-enhanced electrometers and magnetometers that can surpass classical sensitivity limits under controlled conditions. Their role would be to augment classical ISAC, for instance by providing extremely fine-grained measurements for calibration, interference monitoring, or high-precision localization in small regions. Key barriers at this stage include cryogenic or tightly controlled environmental requirements, large size and power consumption of current quantum setups, and the need for robust interfaces between quantum sensing hardware and existing RF/optical front-ends.	
	\item \textit{Phase 2 – Quantum Resources for Distributed Synchronization and Secure Coordination (Beyond Initial 6G Deployments):}
	Once quantum sensing at individual sites matures, a longer-term research direction is to exploit quantum resources such as entanglement for network-level functions. In principle, quantum metrology suggests that entangled states can achieve more favorable scaling of estimation precision with the number of particles than classical schemes, which could be beneficial for distributed timing, frequency synchronization, and cooperative sensing among multiple ISAC nodes. In parallel, quantum key distribution (QKD) and related techniques offer information-theoretic security guarantees that could strengthen ISAC links in highly sensitive applications. However, realizing such capabilities at the scale and robustness required for cellular networks depends on breakthroughs in quantum repeaters, long-distance entanglement distribution, and quantum memories, all of which are currently open research problems.
	\item \textit{Phase 3 – Integrated Quantum-Enabled ISAC Architectures (Post-6G / Beyond IMT-2030):}
	In a far-future horizon, one can envision architectures where quantum-enhanced sensing, quantum-secured communication, and classical ISAC functions are jointly optimized. In such a scenario, quantum resources would be treated as another dimension in the design space, alongside spectrum, power, and time–frequency resources. Achieving this level of integration requires not only hardware breakthroughs (e.g., compact, possibly room-temperature quantum sensors and efficient quantum–classical transducers) but also new theoretical frameworks that unify quantum sensing and communications with classical network optimization and resource allocation. Given the current state of the art, this phase clearly lies beyond the expected 6G commercialization window and should be regarded as a research vision rather than a near-term objective.
\end{itemize}

\textbf{Key Challenges and Theoretical Gaps:} 
Several challenges underline why Quantum-Enabled ISAC is fundamentally more distant than Semantic ISAC. On the hardware side, many of the most advanced quantum platforms still require cryogenic operation, precise isolation from environmental noise, and complex optical or microwave setups that are not yet compatible with cost and form-factor constraints of telecom deployments. On the theory side, while quantum metrology and quantum information provide powerful tools at the link level, a comprehensive system-level theory for quantum-enhanced ISAC in multi-cell, interference-limited, and mobility-constrained networks is still lacking. These gaps justify positioning Quantum-Enabled ISAC as a long-term, post-6G research direction which complements, rather than competes with, the IMT-2030-driven evolution of classical ISAC.

\subsection{Security Challenges in ISAC Systems}
The dual-functional nature of ISAC systems introduces unprecedented security vulnerabilities that extend beyond conventional wireless communication threats, requiring novel defense mechanisms to address cross-domain attack vectors \cite{3.4.2.1}. Physical layer security becomes particularly challenging as ISAC signals must interact fully with their surroundings for effective sensing, creating inherent exposure to eavesdropping where malicious actors can intercept both communication data and sensitive environmental information such as target locations and behavioral patterns \cite{3.1.2.2}. Advanced jamming attacks pose significant threats through sophisticated strategies that can selectively disrupt sensing or communication components, or simultaneously compromise both functions using techniques such as DISCO attacks with illegitimate reconfigurable intelligent surfaces \cite{7.2A.1, 7.2A.2}. Cross-layer vulnerabilities emerge where adversarial attacks on sensing algorithms can manipulate environmental perception while communication-layer threats compromise cooperative sensing scenarios, necessitating artificial intelligence-enabled security schemes and quantum-resistant cryptographic protocols for comprehensive protection \cite{7.2A.3, 7.2.2}. Privacy preservation becomes complex as sensing capabilities can inadvertently collect personally identifiable information through location tracking and behavioral monitoring, requiring development of privacy-preserving sensing algorithms and regulatory compliance frameworks that address the ubiquitous nature of ISAC sensing \cite{7.2A.4}.

\subsection{Evolving Regulatory Landscapes}
As these unified capabilities advance, standards and regulations must adapt to ensure their appropriate utilization \cite{6.2},\cite{7.2.1}. The synthesis of sensing, communication, and localization presents intricate issues related to electromagnetic coexistence, data privacy, and spectrum management \cite{3.3.3},\cite{5.1.1}. Regulators must achieve a delicate balance, facilitating common spectrum accessibility for diverse applications while protecting against unauthorized monitoring and interference \cite{7.2.2}. This is particularly crucial for mission-critical or human-centered deployments where the distinction between sensing and communication is unclear \cite{7.2.3}. Innovative strategies, including adaptive spectrum sharing and context-aware licensing, are becoming prominent, enabling the dynamic utilization of underused frequency bands \cite{5.3.1}. Additionally, innovations in AI provide tools for real-time spectrum coordination and automated compliance enforcement \cite{7.2.4}. Globally, the alignment of regulatory frameworks concerning ethical sensing, security, and privacy will be essential for ensuring interoperability and fostering public confidence \cite{3.4.1},\cite{7.2.5}. Continuous collaboration across public stakeholders, companies, and standardizing organizations is crucial for anticipating disruptive technologies and facilitating their societal integration \cite{4.2.1}.

\subsection{Integration with Emerging Paradigms: NTN, Edge Intelligence, and Cross-Domain Convergence}
To effectively exploit the potential of future networks, unified sensing and communication capabilities must be coordinated with other disruptive innovations. The integration of non-terrestrial networks (NTN), such as high-altitude platforms, unmanned aerial vehicles, and satellite constellations, will provide robust, broadly accessible services \cite{6.3.5},\cite{7.3.1}. These hybrid architectures are essential for providing smart connectivity to disaster-affected, underserved, or remote regions where terrestrial networks are inadequate \cite{7.3.2}. Simultaneously, integrating lightweight AI at the network edge enables devices to make localized decisions that collectively enhance data transmission and environmental awareness. This enables ultra-low latency services while enhancing autonomy and maintaining user privacy  \cite{6.3.7}. Broader integration with sectors, including smart energy grids, cyber-physical infrastructure, and robotics, facilitates real-time coordination across several layers of functioning \cite{3.4.2},\cite{7.3.3}. Technologies such as collaborative control systems and federated learning facilitate the development of secure, sustainable, and adaptable infrastructures that respond to both physical and digital signals \cite{3.4.1}\cite{7.3.4}. These advancements require flexible, composable designs that enhance interoperability and can scale among several sectors, establishing a foundation for extremely responsive, future-oriented wireless ecosystems \cite{6.3.11},\cite{7.3.5},\cite{7.3.6}. Model-based systems engineering (MBSE) 2.0 frameworks are expected to play a central role in this transformation, providing the tools needed for integrated modeling, validation, and intelligent control of complex, multi-domain 6G networks \cite{7.3.7}.

\section{Conclusion} \label{conclude}
ISAC stands as a transformative cornerstone for next-generation 6G networks, merging sensing and communication into unified signal processing, waveforms, and hardware systems to deliver substantial gains in cost, energy, and spectrum efficiency. This tutorial surveyed ISAC's lineage from radar-communication coexistence to dual-functional transceivers, analyzed core enabling technologies such as massive MIMO, RISs, and AI-driven architectures, and highlighted ISAC's potential to revolutionize wireless networks into adaptive, intelligent systems through real-time resource management in diverse applications like extended reality and intelligent transportation. However, realizing ISAC’s full promise requires overcoming significant research and standardization challenges, including development of unified performance metrics, advanced protocols, scalable hardware, robust interference management, and privacy solutions for human-centric scenarios, demands that necessitate close collaboration among academia, industry, and standards bodies. As 6G matures, ISAC is poised to underpin advancements in semantic communication, quantum-enhanced environmental awareness, and networked sensing, with future progress in IoT, edge intelligence, NTNs, and adaptive AI-driven hardware paving the way for networks where seamless interaction among humans, machines, and environments fosters a new era of connectivity and pervasive intelligence.


%



\section*{Acknowledgment}
This work is partially supported by NSF  ECCS-2302469,  and Japan Science and Technology Agency (JST) Adopting Sustainable Partnerships for Innovative Research Ecosystem (ASPIRE) JPMJAP2326. The work of O. Günlü was partially supported by the ZENITH Research and Leadership Career Development Fund under Grant ID23.01, EU COST Action 6G-PHYSEC, Swedish Foundation for Strategic Research (SSF) under Grant ID24-0087, and German Federal Ministry of Research, Technology and Space (BMFTR) 6GEM+ Transfer Hub under Grants 16KIS2412 and 16KISS005. T. Riihonen's research work was supported by the Research Council of Finland (grant \#341489). The work of Yuanhao Cui was supported in part by China Association for Science and Technology (CAST) Young Talent Support Program under Grant No. YESS2023066. The work of M. Di Renzo was supported in part by the European Union through the Horizon Europe project COVER under grant agreement number 101086228, the Horizon Europe project UNITE under grant agreement number 101129618, the Horizon Europe project INSTINCT under grant agreement number 101139161, and the Horizon Europe project TWIN6G under grant agreement number 101182794, as well as by the Agence Nationale de la Recherche (ANR) through the France 2030 project ANR-PEPR Networks of the Future under grant agreement NF-YACARI 22-PEFT-0005, and by the CHIST-ERA project PASSIONATE under grant agreements CHIST-ERA-22-WAI-04 and ANR-23-CHR4-0003-01. Also, the work of M. Di Renzo was supported in part by the Engineering and Physical Sciences Research Council (EPSRC), part of UK Research and Innovation, and the UK Department of Science, Innovation and Technology through the CHEDDAR Telecom Hub under grant EP/X040518/1 and grant EP/Y037421/1, and through the HASC Telecom Hub under grant EP/X040569/1.


\ifCLASSOPTIONcaptionsoff
  \newpage
\fi



%

\bibliographystyle{IEEEtran}
\bibliography{sample-base}

%

\begin{IEEEbiography}[{\includegraphics[width=1in,height=1.25in,clip,keepaspectratio]{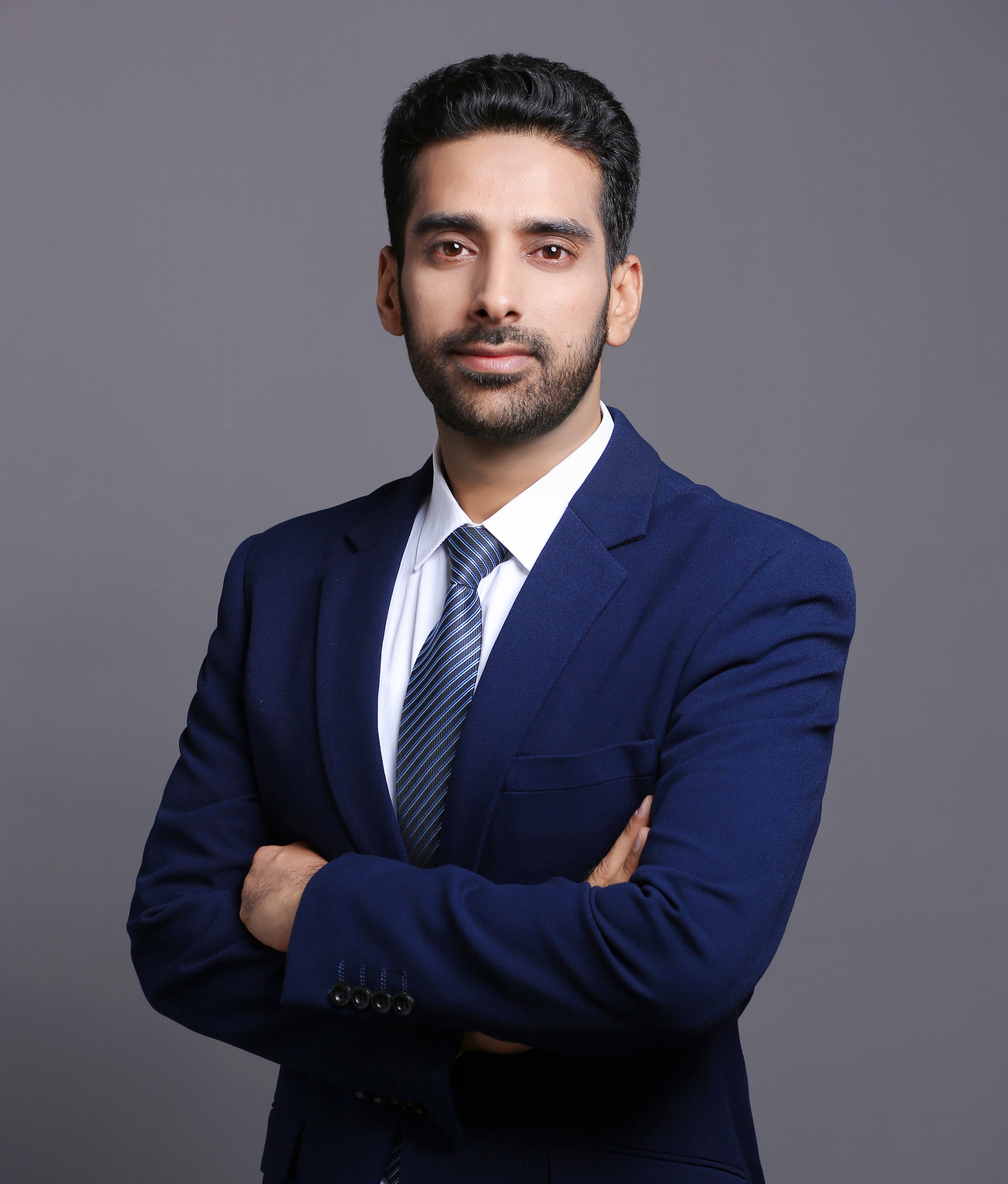}}]{Muhammad Umar Farooq Qaisar}
	(Member, IEEE) received the B.S. degree in Computer Science from the International Islamic University, Islamabad, Pakistan, in 2012, and the M.S. degree in Computer Science and Technology from the University of Science and Technology of China (USTC) in 2017. He earned the Ph.D. degree in Computer Science and Technology from USTC in 2022. He was a postdoctoral fellow with the School of System Design and Intelligent Manufacturing, Southern University of Science and Technology, Shenzhen, China. He later served as an Associate Professor in the School of Computer Science, Northwestern Polytechnical University, Xi’an, China. He is currently an Associate Professor with the Hangzhou International Innovation Institute, Beihang University, Hangzhou, China. His research interests include the Internet of Things (IoT), wireless sensor networks (WSN), wireless rechargeable sensor networks (WRSN), software-defined networking (SDN), vehicular ad hoc networks (VANETs), integrated sensing and communication (ISAC), unmanned aerial vehicles (UAVs), and communication security.
\end{IEEEbiography}
\vskip -2\baselineskip plus -1fil
\begin{IEEEbiography}[{\includegraphics[width=1in,height=1.25in,clip,keepaspectratio]{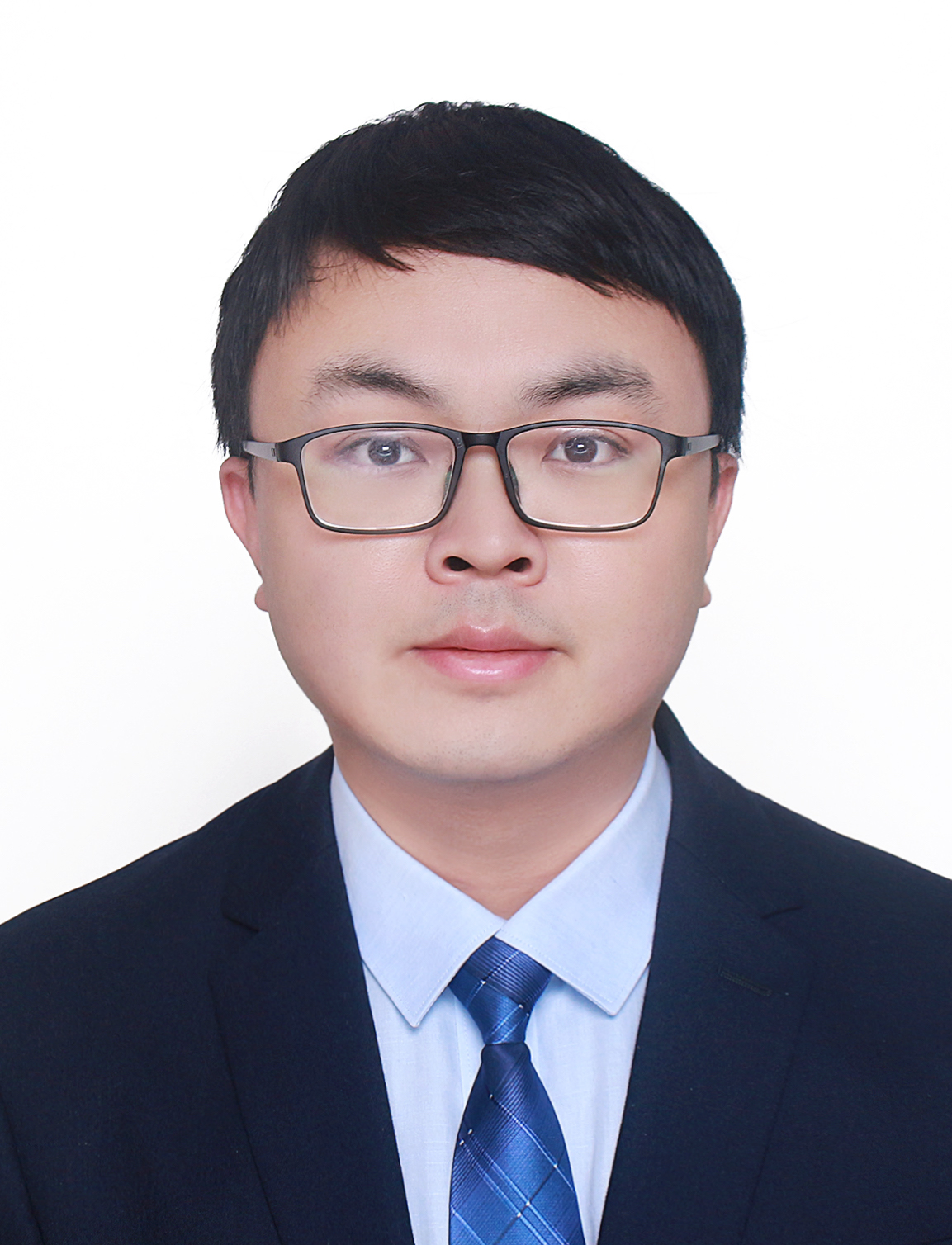}}]{Weijie Yuan}
	(Senior Member, IEEE) ‘s research interests include Integrated Sensing and Communications (ISAC), Orthogonal Time Frequency Space (OTFS), and the Low-Altitude Wireless Networks (LAWN). He currently serves as the Deputy Editor-in-Chief of the Journal of Advances in Signal Processing. He is an Editor for the IEEE Transactions on Communications, IEEE Transactions on Wireless Communications, IEEE Transactions on Mobile Computing, IEEE Communications Magazine, IEEE Communications Standards Magazine, IEEE Transactions on Green Communications and Networking, IEEE Communications Letters, IEEE Open Journal of Communications Society, and npj Wireless Technology. He has led four special issues in IEEE Transactions on Vehicular Technology, IEEE Transactions on Network Science and Engineering, IEEE Journal of Selected Topics in Signal Processing, and China Communications. He was a Guest Editor for IEEE Internet of Things Journal and IEEE Open Journal of Vehicular Technology. He is serving/served as the the General Co-Chair for ISWCS 2026, Symposium Co-Chair for IEEE/CIC ICCC 2026, and Track Co-Chair for IEEE ICC 2025 and IEEE VTC 2025-Spring. He served as an Organizer/the Chair of several workshops and special sessions in flagship IEEE and ACM conferences, including IEEE ICC, IEEE VTC, IEEE GlobeCom, IEEE/CIC ICCC, IEEE SPAWC, IEEE WCNC, IEEE ICASSP, and ACM MobiCom. He is the Founding Chair of the IEEE Aerospace and Electronic System Working Group on LAWN and the ComSoc Special Interest Group (SIG) on LAWN. He was a recipient of the Best Editor from IEEE CommL, the Best Paper Award from IEEE ICC 2023, IEEE/CIC ICCC 2023, IEEE GlobeCom 2024, and IEEE GlobeCom 2025, as well as the 2025 IEEE Communications Society \& Information Theory Society Joint Paper Award. 
\end{IEEEbiography}

\vskip -2\baselineskip plus -1fil
\begin{IEEEbiography}[{\includegraphics[width=1in,height=1.25in,clip,keepaspectratio]{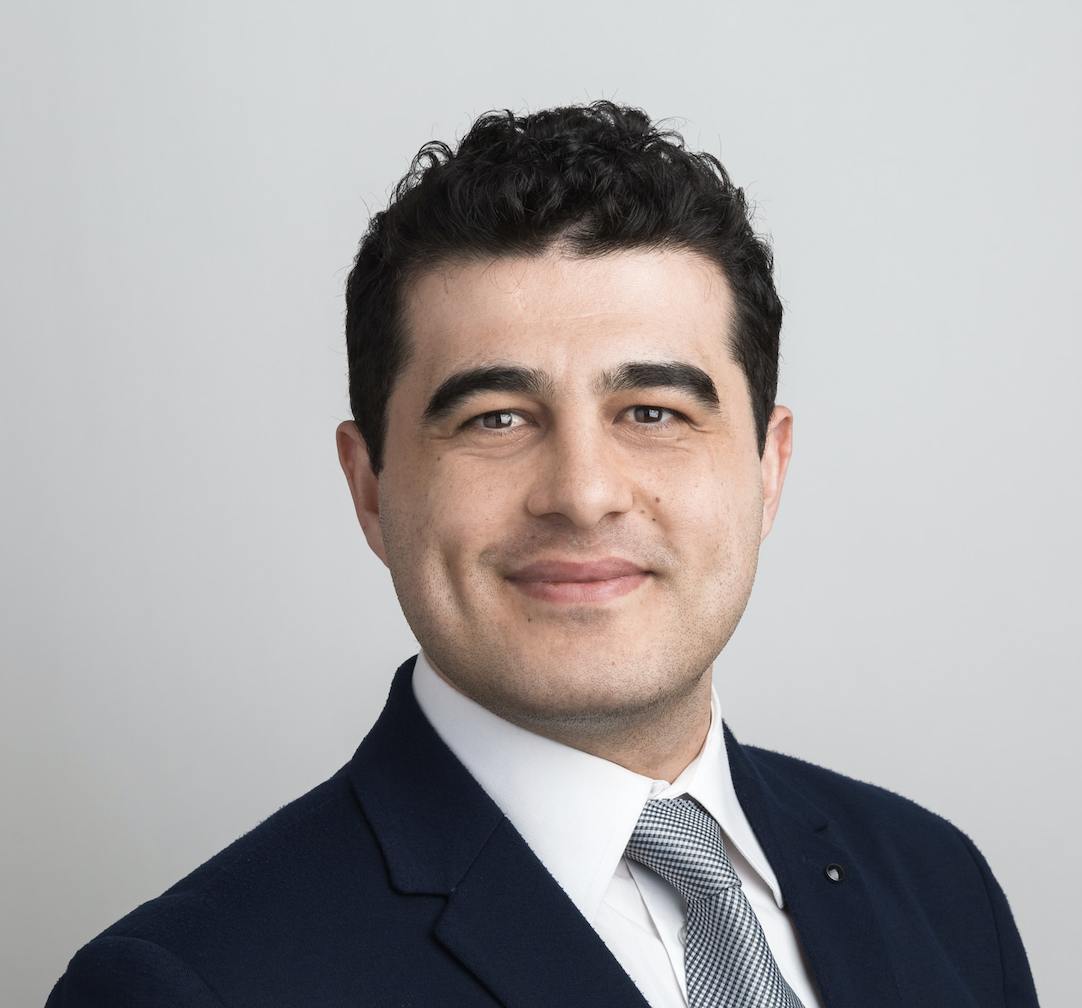}}]{Onur Günlü} (Senior Member, IEEE) received the B.Sc. degree (Highest Distinction) in Electrical and Electronics Engineering from Bilkent University, Turkey in 2011; M.Sc. (Highest Distinction) and Dr.-Ing. (Ph.D. equivalent) degrees in Communications Engineering both from the Technical University of Munich (TUM), Germany in 2013 and 2018, respectively. He was a Working Student in the Communication Systems division of Intel Mobile Communications (IMC), now Apple Inc., in Munich, Germany during November 2012 - March 2013. Onur worked as a Research and Teaching Assistant at TUM Chair of Communications Engineering (LNT) between February 2014 - May 2019. As a Visiting Researcher, among more than twenty Research Stays at Top Universities and Companies, he was at TU Eindhoven, Netherlands during February 2018 - March 2018. Onur was a Visiting Research Group Leader at Georgia Institute of Technology, Atlanta, USA during February 2022 - March 2022. He was also a Visiting Professor at TU Dresden, Germany during February 2023 - March 2023. Following Research Associate and Group Leader positions at TUM, TU Berlin, and the University of Siegen, he joined Linköping University in October 2022 as an ELLIIT Assistant Professor and obtained tenure as an Associate Professor leading the Information Theory and Security Laboratory (ITSL) in August 2024. He became a Swedish Docent (Habilitation) of Information Theory in December 2023 and an IEEE Senior Member in July 2024. Since September 2025, Onur has been a Tenured Full Professor leading the Lehrstuhl für Nachrichtentechnik (Institute of Communications Engineering ) at TU Dortmund (TUDO), Germany and a Guest Professor at Linköping University, Sweden. 
	
Onur has received the 2025 IEEE Information Theory Society - Joy Thomas Tutorial Paper Award, the 2023 ZENITH Research and Career Development Award, 2021 IEEE Transactions on Communications - Exemplary Reviewer Award, and the prestigious VDE Information Technology Society (ITG) 2021 Johann-Philipp-Reis Award. His research interests include distributed function computation, information-theoretic privacy and security, coding theory, integrated sensing and communication, and private learning. Among his publications is the book \emph{Key Agreement with Physical Unclonable Functions and Biometric Identifiers} (Dr. Hut Verlag, 2019). He serves as Associate Editor for {IEEE JOURNAL ON SELECTED AREAS IN COMMUNICATIONS}, {IEEE TRANSACTIONS ON COMMUNICATIONS}, and {ENTROPY} Journal, and was recently an Associate Editor of {EURASIP JOURNAL ON WIRELESS COMMUNICATIONS AND NETWORKING} and a Guest Editor of {IEEE JOURNAL ON SELECTED AREAS IN INFORMATION THEORY}. He also serves as a Board Member and Secretary of the IEEE Sweden VT/COM/IT Joint Chapter, and as a Working Group Leader for EU COST Action 6G-PHYSEC on Intelligent and Resilient Systems.
\end{IEEEbiography}
\vskip 0\baselineskip plus -1fil
\vspace{-11 mm}
\begin{IEEEbiography}[{\includegraphics[width=1in,height=1.25in,clip,keepaspectratio]{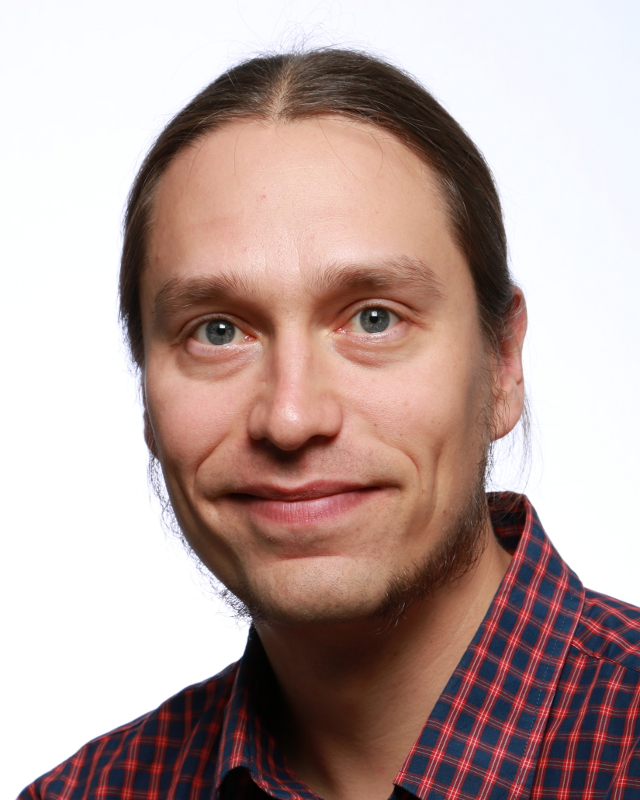}}]{Taneli Riihonen} (Senior Member, IEEE) received his D.Sc. degree in electrical engineering (with honors) from Aalto University, Helsinki, Finland, in 2014 and completed his postdoctoral phase at Columbia University, New York, USA. He is now a tenure-track Associate Professor at Tampere University, Finland. He received the Finnish technical sector's award for the best doctoral dissertation of the year and the EURASIP Best PhD Thesis Award 2017 as well as won the EDA Defence Innovation Prize 2020. He has served as an Editor for IEEE Communications Letters, IEEE Wireless Communications Letters, and IEEE Transactions on Wireless Communications. His research is focused on physical-layer analysis, link-layer techniques and signal processing for all kinds of radio systems from consumer and commercial domains to defense and security with current interest in the evolution of 6G.
\end{IEEEbiography}
\vskip 0\baselineskip plus -1fil
\vspace{-11 mm}
\begin{IEEEbiography}[{\includegraphics[width=1in,height=1.25in,clip,keepaspectratio]{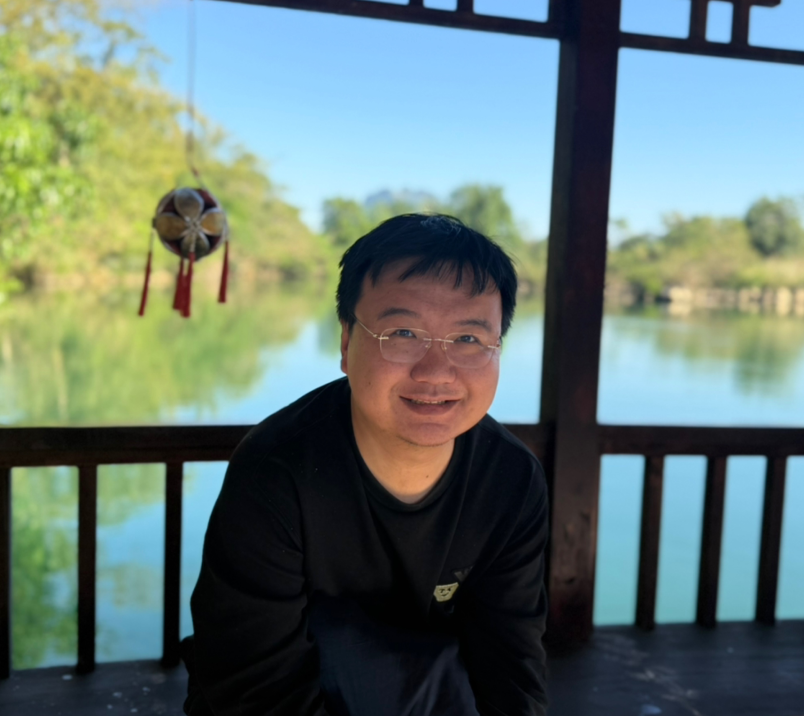}}]
{Yuanhao Cui} (Member, IEEE) is currently an assistant Professor with the School of Information Science and Engineering, Beijing University of Posts and Telecommunications, Beijing, China. Dr. Cui’s research interests lie in the general area of signal processing and wireless communications, and in particular in the area of Integrated Sensing and Communications (ISAC) and Low-Altitude Wireless Network (LAWN). He is the Founding Chair of the IEEE ComSoc Special Interest Group on Low-Altitude Wireless Networks (LAWN-SIG), the founding Secretary of the IEEE ComSoc ISAC Emerging Technology Initiative (ISAC-ETI), and the founding Secretary of the CCF Scientific Communication standing committee. He serves on the editorial board of IEEE Transactions on Mobile Computing, IEEE Vehicular Technology Magazine, IEEE Journal of Internet of Things, and IEEE Journal of Biomedical and Health Informatics. He is a member of the IMT-2030 (6G) ISAC Task Group. He was listed among the World’s Top 2\% Scientists by Stanford University for citation impact from 2023 to 2025. He was a recipient of numerous Best Paper Awards, including the 2025 IEEE Communication Society and Information Theory Society Joint Paper Award, 2024 IEEE Communications Society Asia-Pacific Outstanding Paper Award, 2024 IEEE Globecom Best Paper Award, 2024 IEEE JC\&S Symposium Best Paper Award, 2023 ACM MobiCom Best Paper Award in ISAC, and 2023 IEEE/CIC ICCC 2023 Best Paper Award.
\end{IEEEbiography}
\vskip 0\baselineskip plus -1fil
\vspace{-11 mm}
\begin{IEEEbiography}[{\includegraphics[width=1in,height=1.25in,clip,keepaspectratio]{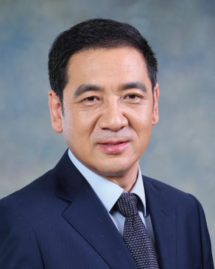}}]{Lin Zhang} (Senior Member, IEEE) received the B.S. degree from the Department of Computer and System Science, Nankai University, Tianjin, China, in 1986, and the M.S. and Ph.D. degrees from the Department of Automation, Tsinghua University, Beijing, China, in 1989 and 1992, respectively. From 2002 to 2005, he worked as a Senior Research Associate at the U.S. Naval Postgraduate School, U.S. National Research Council. He is currently a Professor at Beihang University. He is a Chief Scientist of the National 863 Program and the National Key Research and Development Program of China. He serves as the Director of the Engineering Research Center of Complex Product Advanced Manufacturing Systems, Ministry of Education of China. He has authored or coauthored more than 300 papers and 20 books and chapters. His research interests include service-oriented modeling and simulation, cloud manufacturing and simulation, model engineering, model-based systems engineering, cyber-physical systems, and modeling and simulation for manufacturing systems.
	
Dr. Zhang is a fellow of SCS, ASIASIM, and CSF. He served as the President of the Society for Modeling and Simulation International (SCS), the Executive Vice President of the China Simulation Federation (CSF), and the President of the Federation of Asian Simulation Societies (ASIASIM). He received the National Award for Excellent Science and Technology Books, the Outstanding Individual Award of the National High-Tech Research and Development Program, the National Excellent Scientific and Technological Workers Awards, and the SCS Outstanding Professional Contribution Award. He serves as an Editor-in-Chief and Associate Editor of eight peer-reviewed international journals.	
\end{IEEEbiography}
\vspace{-13 mm}
\vskip 0\baselineskip plus -1fil
\begin{IEEEbiography}[{\includegraphics[width=1in,height=1.25in,clip,keepaspectratio]{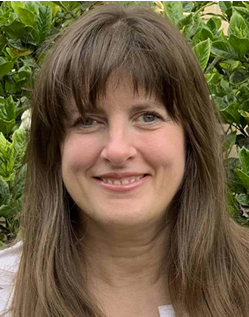}}]{Nuria Gonzalez-Prelcic} (Fellow, IEEE) received the Ph.D. with Honors in telecommunication engineering in 2000 from the University of Vigo, Spain. She is currently a Professor at the ECE Department of the University of California San Diego. She previously held faculty positions at the ECE Department of NC State University, USA (2020–2023) and at the Signal Theory and Communications Department at the University of Vigo, Spain (2000–2020). She also held visiting positions at the University of Texas at Austin and the University of New Mexico. She was also the founding director of the Atlantic Research Center for Information and Communication Technologies (atlanTTic) at the University of Vigo (2008–2017). She is currently a Distinguished Lecturer in the IEEE Vehicular Technology Society. Her main research interests include signal processing and machine learning for wireless communications and sensing, with a focus on MIMOprocessing for mmWave communication, joint sensing and communication, sensor-aided communication, signal processing under hardware impairments, vehicular communication, and multiantenna technology for LEO satellite communication. She has published more than 150 papers in these areas, including a highly cited tutorial on signal processing for mmWave MIMO published in the IEEE Journal of Selected Topics in Signal Processing which has received the 2020 IEEE SPS Donald G. Fink Overview Paper Award, and a paper pioneering the idea of enabling automotive radar with a WiFi waveform that won the 2022 IEEE Vehicular Technology Society Best Vehicular Electronics Paper Award. Her work on self-attention networks for user position tracking received a best student paper award at the 2023 IEEE Signal Processing for Wireless Communications (SPAWC) conference. She was an Editor for IEEE Transactions on Wireless Communications and IEEE Transactions Communications. She is a member of the IEEE Signal Processing Society TWG on Integrated Sensing and Communication, SPCOM Technical Committee, and IEEE SPS Education Board. 
\end{IEEEbiography}

\vspace{-13 mm}
\vskip 1\baselineskip plus -1fil
\begin{IEEEbiography}[{\includegraphics[width=1in,height=1.25in,clip,keepaspectratio]{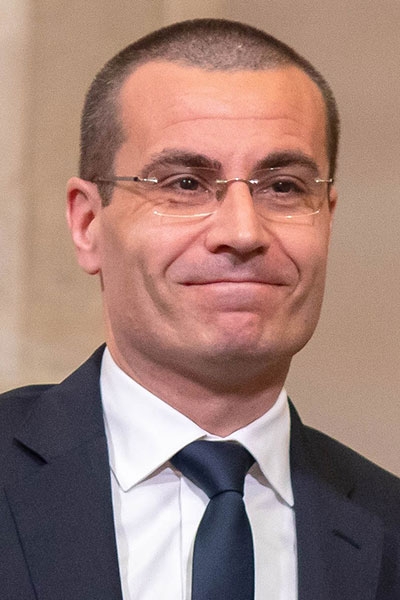}}]{Marco Di Renzo} (Fellow, IEEE) received the Laurea (cum laude) and Ph.D. degrees in electrical engineering from the University of L’Aquila, Italy, in 2003 and 2007, respectively, and the Habilitation à Diriger des Recherches (Doctor of Science) degree from University Paris-Sud (currently Paris-Saclay University), France, in 2013. Currently, he is Chair Professor of Telecommunications Engineering, the Director of the Centre for Telecommunications Research, and the Head of the Telecommunications Group, Department of Engineering, King’s College London, London, United Kingdom. He is also a CNRS Research Director (Professor) with the Laboratory of Signals and Systems at CNRS-CentraleSupélec, Paris-Saclay University, Paris, France. He was a France-Nokia Chair of Excellence in ICT at the University of Oulu (Finland), a Tan Chin Tuan Exchange Fellow in Engineering at Nanyang Technological University (Singapore), a Fulbright Fellow at The City University of New York (USA), a Nokia Foundation Visiting Professor at Aalto University (Finland), and a Royal Academy of Engineering Distinguished Visiting Fellow at Queen’s University Belfast (U.K.). He is a Fellow of the IEEE, IET, EURASIP, and AAIA; an Academician of AIIA; an Ordinary Member of the European Academy of Sciences and Arts, an Ordinary Member of the Academia Europaea, and an Ordinary Member of the Italian Academy of Technology and Engineering; an Ambassador of the European Association on Antennas and Propagation; and a Highly Cited Researcher. He has received several distinctions, including the Michel Monpetit Prize conferred by the French Academy of Sciences, the IEEE Communications Society Heinrich Hertz Award, and the IEEE Communications Society Marconi Prize Paper Award in Wireless Communications. Also, he is a principal investigator of an ERC Synergy grant on metasurface-based information processing. He served as the Editor-in-Chief of IEEE Communications Letters from 2019 to 2023, and as the Director of Journals and Chair of the Publications Misconduct Ad Hoc Committee of the IEEE Communications Society from 2024 to 2025. Currently, he sits on the IEEE-COMSOC Fellow Evaluation Standing Committee and on the Editorial Board of the Proceedings of the IEEE. 
\end{IEEEbiography}
\vspace{-13 mm}
\vskip 0\baselineskip plus -1fil
\begin{IEEEbiography}[{\includegraphics[width=1in,height=1.25in,clip,keepaspectratio]{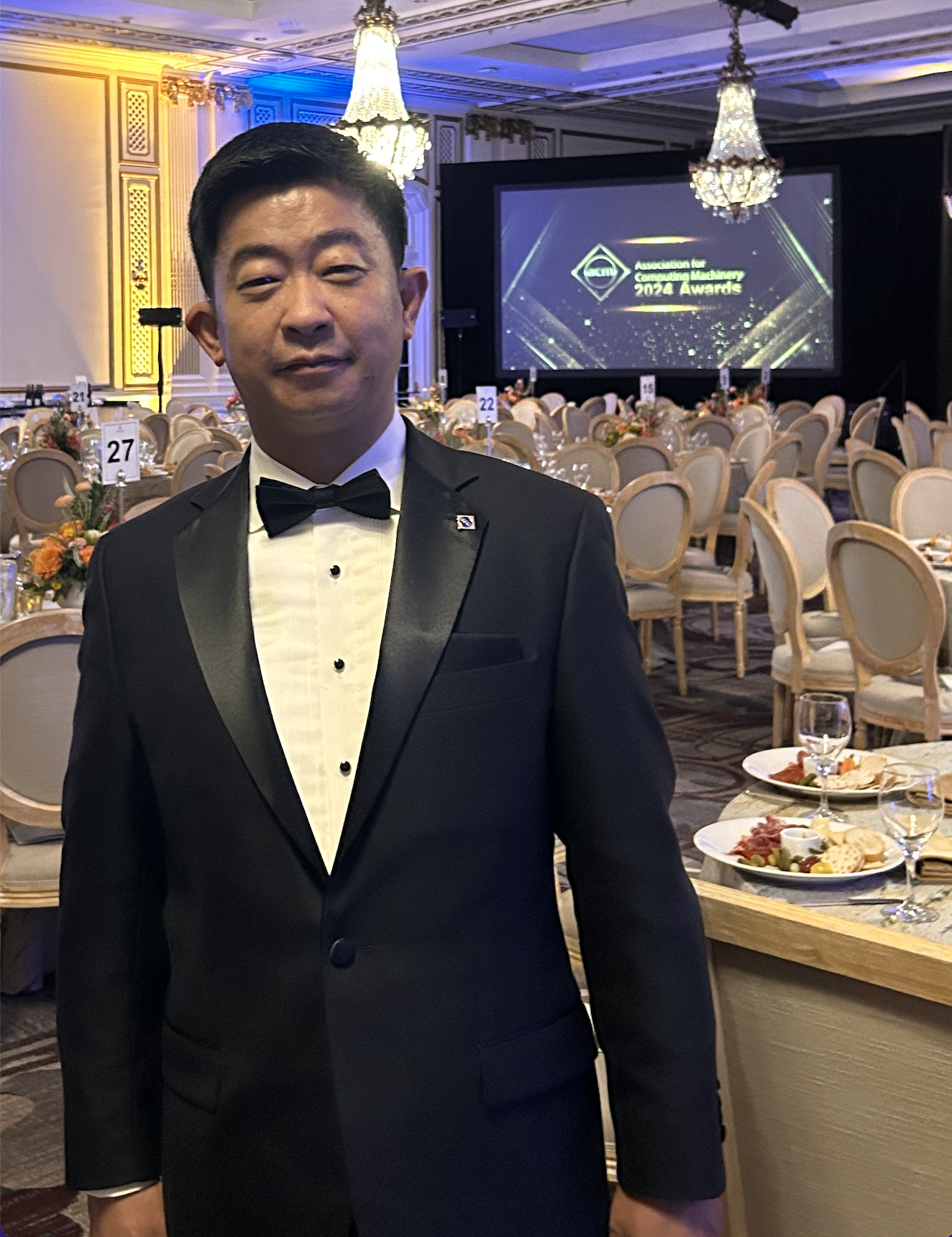}}]{Zhu Han} (Fellow, IEEE) (S’01–M’04-SM’09-F’14) received the B.S. degree in electronic engineering from Tsinghua University, in 1997, and the M.S. and Ph.D. degrees in electrical and computer engineering from the University of Maryland, College Park, in 1999 and 2003, respectively. 
	
From 2000 to 2002, he was an R\&D Engineer of JDSU, Germantown, Maryland. From 2003 to 2006, he was a Research Associate at the University of Maryland. From 2006 to 2008, he was an assistant professor at Boise State University, Idaho. Currently, he is a John and Rebecca Moores Professor in the Electrical and Computer Engineering Department as well as in the Computer Science Department at the University of Houston, Texas. Dr. Han’s main research targets on the novel game-theory related concepts critical to enabling efficient and distributive use of wireless networks with limited resources. His other research interests include wireless resource allocation and management, wireless communications and networking, quantum computing, data science, smart grid, carbon neutralization, security and privacy.  Dr. Han received an NSF Career Award in 2010, the Fred W. Ellersick Prize of the IEEE Communication Society in 2011, the EURASIP Best Paper Award for the Journal on Advances in Signal Processing in 2015, IEEE Leonard G. Abraham Prize in the field of Communications Systems (best paper award in IEEE JSAC) in 2016, IEEE Vehicular Technology Society 2022 Best Land Transportation Paper Award, and several best paper awards in IEEE conferences. Dr. Han was an IEEE Communications Society Distinguished Lecturer from 2015 to 2018 and ACM Distinguished Speaker from 2022 to 2025, AAAS fellow since 2019, and ACM Fellow since 2024. Dr. Han is also the winner of the 2021 IEEE Kiyo Tomiyasu Award (an IEEE Field Award), for outstanding early to mid-career contributions to technologies holding the promise of innovative applications, with the following citation: ``for contributions to game theory and distributed management of autonomous communication networks." Dr. Han is honored Lifetime Chair Professor of National Yang Ming Chiao Tung University, Taiwan, Eminent Scholar of Kyung Hee University, South Korea and Global Professor of Keio University, Japan.  
\end{IEEEbiography}




\end{document}